\documentclass[fleqn,10pt]{wlscirep}
\usepackage[utf8]{inputenc}
\usepackage[T1]{fontenc}

\usepackage{setspace}
\usepackage{graphicx}
\usepackage{rotating}
\usepackage{csquotes}
\usepackage{xcolor}
\usepackage{nameref}

\newcommand{\hl}[1]{\textcolor{black}{#1}} 

\newcommand{\argmax}{\operatornamewithlimits{argmax}}
\newcommand{\mty}[1]{\text{\small{#1}}} 

\title{Loan maturity aggregation in interbank lending networks obscures mesoscale structure and economic functions}

\author[1]{Marnix Van Soom}
\author[2,3,*]{Milan van den Heuvel}
\author[2]{Jan Ryckebusch}
\author[3,4]{Koen Schoors}
\affil[1]{Vrije Universiteit Brussel, Artificial Intelligence Lab, Brussels, 1050, Belgium}
\affil[2]{Ghent University, Department of Physics and Astronomy, Ghent, 9000, Belgium}
\affil[3]{Ghent University, Department of Economics, Ghent, 9000, Belgium}
\affil[4]{National Research University, Higher School of Economics, Moscow, Russia}
\affil[*]{Milan.vandenHeuvel@UGent.be}


\begin{abstract}
Since the 2007-2009 financial crisis, substantial academic effort has been dedicated to improving our understanding of interbank lending networks (ILNs). Because of data limitations or by choice, the literature largely lacks multiple loan maturities. We employ a complete interbank loan contract dataset to investigate whether maturity details are informative of the network structure. Applying the layered stochastic block model of Peixoto (2015) and other tools from network science on a time series of bilateral loans with multiple maturity layers in the Russian ILN, we find that collapsing all such layers consistently obscures mesoscale structure. The optimal maturity granularity lies between completely collapsing and completely separating the maturity layers and depends on the development phase of the interbank market, with a more developed market requiring more layers for optimal description. Closer inspection of the inferred maturity bins associated with the optimal maturity granularity reveals specific economic functions, from liquidity intermediation to financing. Collapsing a network with multiple underlying maturity layers or extracting one such layer, common in economic research, is therefore not only an incomplete representation of the ILN’s mesoscale structure, but also conceals existing economic functions. This holds important insights and opportunities for theoretical and empirical studies on interbank market functioning, contagion, stability, and on the desirable level of regulatory data disclosure.

\end{abstract}
\begin{document}

\flushbottom
\maketitle
\thispagestyle{empty}

\section*{Introduction and overview}

Interbank lending networks (ILNs) are complex network models of the interbank money markets, often called the plumbing of modern financial systems\cite{bargigli2015}. Banks make interbank loans on such markets to accommodate daily liquidity imbalances \hl{and manage their duration gap, defined as the maturity disparity between bank assets and liabilities~\cite{Bluhm2016}}. For example, a bank holding more cash than desired may profitably lend this cash to other banks in need of cash~\cite{wiemers2003}.

The financial crisis of 2007-2009 has brought the interbank money markets in the public eye because their ``drying up'' (failure to lend cash to banks in need) was a major channel of financial contagion~\cite{gai2011}. Since then, substantial academic effort has been dedicated towards improving our understanding of these markets. It turns out that representing the interbank money market as a network is a simple and powerful abstraction~\cite{Caccioli2018}, that is arguably more realistic than modelling it as one representative bank, as is customary in traditional macro-finance~\cite{Huser2015}. Network analysis \hl{is} therefore now one of the standard tools of financial stability experts worldwide, i.a. at the IMF\cite{Chan2009} and ECB~\cite{Tumpell2009}.

A thorough understanding of the interbank money market is vital to prevent systemic meltdowns~\cite{Battiston2016, Roukny2018}, implying a need for ever more realistic models of ILNs. We contribute by studying a much neglected aspect of ILNs, i.e.~the different loan maturities. The maturity of a loan, i.e.~the period after which the loan must be repaid, is an important instrument for banks to organise their lending and borrowing activity in function of risk minimisation~\cite{Rochet1996,Furfine2001, Bluhm2016}. A majority of empirical financial network papers study the overnight interbank market~\cite{Huser2015}. \hl{This seems to partly originate from the widely held view that overnight lending makes up the majority of interbank exposure, a view that stands in contrast with recent results showing the average contract length in the German interbank market to be well over one year~\cite{Bluhm2016}. Sometimes however, empirical ILN literature is unable to include different loan maturities because of data limitations (see~\cite{Anand2018} for a recent overview of available interbank data). As a consequence} agent-based models of ILNs often either neglect loan maturities or limit it to a modelling detail~\cite{halaj2015,xu-model2016}, even though maturity choices reflect bank risk strategies. For example, stress testing models would benefit from exposure data enriched with maturity information~\cite{Huser2015}. Enabled by a particularly granular dataset, a panel of all lending contracts issued in the Russian ILN~\cite{Karas2010}, we investigate qualitatively what kind of information is lost by not differentiating between the loan maturities. We further try to estimate the pitfalls of this common practice. The approach we take consists of explicitly modelling the Russian ILN by layered stochastic block models (SBMs). This allows us to determine to what extent the loan maturities are informative of the Russian ILN's \hl{\emph{mesoscale}}, i.e.~the higher-level organisation of the banks into bank groups.

The dataset used in this work consists of 57 monthly ILNs constructed from the complete panel of contracts on the Russian domestic unsecured interbank lending market, one for each month in the period from January 2000 to October 2004, except for January 2003. There are a total of 2.4 million loans, each annotated with its lender, borrower, month of issuance, loan size and \emph{maturity class}. \hl{The dataset is unusual in the sense that it is an loan issuance network rather than an exposure network, the latter being the typical ILN representation found in the literature~\cite{Anand2018}. The exposure network associated with a given issuance network can be derived by aggregating the loan sizes in the issuance network over time.} \hl{The data originates directly from the bank reports to the Central Bank of Russia (CBR) and came preassigned into eight contiguous maturity classes, so that every loan can be assigned to a particular maturity class~(e.g.~2-7 days).}
Fig.~\ref{phase-loans} shows the monthly number of active banks, loans issued, and outstanding loans per maturity class, together with a list of the eight maturity classes (further details on the dataset can be found in the Methods section). We look at the monthly ILNs as layered networks such that each maturity class corresponds to one \emph{maturity layer}. By putting contiguous maturity layers into bins, they can be coarse-grained to achieve various levels of \emph{layer granularity}, ranging from complete differentiation (eight bins, one layer per bin) to complete aggregation (all layers merged into one bin, the collapsed network). 

First, we provide descriptive statistics of the monthly ILNs in a comparative framework. We characterise the network topology of each maturity layer separately using the typical network measures from ILN literature (listed in the leftmost column of Table~\ref{stylized}) and compare the results across the maturity layers and across the literature. Broadly speaking and integrating over time, we find layer non-homogeneity for most topology measures. This means that, while stylised facts found in the literature were also found in some of the maturity layers, the topology measures do not take on similar values for all maturity layers simultaneously. Layer non-homogeneity points to the fact that complete aggregation or single layer focus involves the loss of the topological diversity present in the maturity layers, although some layers do share similar lending patterns in specific months (by lending pattern we mean a topological pattern in an ILN). For example the three most dense layers -- the ``short'' layers, i.e.~all maturities below 30 days -- display a strong similarity throughout the full time period. The short layers' topological similarity, together with their dominant share in the number of loans issued, could suggest that complete aggregation might not be harmful after all, irrespective of what the found layer non-homogeneity suggests. The question then becomes whether the information lost by complete aggregation is relevant for ILN structure and economic function. Because financial stability plays a central role in literature and policy, we are interested in the money markets' mesoscale organisation, which can effectuate the propagation of instability and risk~\cite{Roukny2013}, rather than individual banks' lending strategies. Relevant information is thus any set of information that allows for characterisation of the mesoscale structure rather than other, more specific, more ``noisy'', information about local lending patterns (e.g.~clustering).

\hl{In order to characterize the effect of aggregating maturity layers on the ILN mesoscale, we explicitly model the monthly ILNs by layered SBMs. SBMs can infer statistically significant group structure in networks, without making informative prior assumptions about the type of mesostructure itself. SBMs model the mesoscale of a network by assuming that the nodes in a network behave ``group-like'' rather than on individual account. \emph{Layered} SBMs generalize SBMs for layered networks by allowing the group structure to have a different topological pattern on each layer.} As a concrete example, imagine a network with two layers representing the mating (layer 1) and conflict (layer 2) interactions in a population of deer (the nodes). One possible layered SBM of this network is the division of the deer into two groups, male and female, so that observed occurrences of mating and fighting between two deer are explained only by their sexes. Note that such a simple model -- which does not take things like social status etc. into account to explain the observed interactions -- could well suffice to infer the deer's sexes if these were unknown.

Layered SBMs formulated in a Bayesian setting can be extended to the \emph{coarse-grained} layered SBM in order to infer the appropriate level of layer granularity (the optimal granularity, OG) along with the group structure. The bins in the OG correspond to lending patterns between the bank groups that differ from each other in a statistically significant way. We explain this in detail in the next section \hl{and give a simple illustration in Fig.\ref{wrap}}, but the essence of how an OG can be inferred can already be understood by a simple regularisation argument. The complexity and modelling power of a \hl{coarse-grained} layered SBM is determined \hl{primarily} by the number of groups and the degree of layer granularity, as these parameters simultaneously define the ``resolution'' available to model the observed layered network. To prevent overfitting (i.e.~modelling noise), any increase in model complexity should be warranted by enough statistical evidence in the data. Thus the OG may be determined in general by any regularisation principle to balance model complexity and quality of fit; we use \hl{Bayesian model selection} for this. One specific advantage of this approach is that there is formally no difference between inferring the number of groups -- and in fact the groups themselves -- and the OG; both are determined as part of a single inference for each monthly ILN.

The inferred monthly OGs are displayed in Fig.~\ref{OLP}(a). Each OG consists of a set of bins numbered from short to long maturity by the OG bin index (OGB index). We mention here two findings easily deduced from Fig.~\ref{OLP}(a). First, we find that the OG always lies between complete differentiation and complete aggregation. This means that on the one hand the eight available maturity classes are partly redundant and that the lending patterns may be described more effectively by merging maturity layers into bins, as is indeed the case for the short layers mentioned before\hl{.} On the other hand complete aggregation \hl{apparently} discards important information needed to model the monthly ILN's mesostructure: \hl{The lending patterns between the bank groups depend significantly on the maturity classes of the loans. This also illustrates the more general added benefit of statistically inferring the optimal maturity bins, instead of setting them manually based on some heuristic or prior: if one would set bin widths too narrowly, the data might be too thinly spread to detect any structure at all. Conversely, if one would set bin widths too widely, the different structures, and their economic functions, might be obfuscated.} Second, the monthly \hl{\emph{number of bins}} in the OG correlate roughly with the known two phases in the Russian money market's development: early development (roughly before month 35) and emerging maturity (from month 35 onward) (see Methods). This points to a natural ordering in number and complexity of lending patterns emerging at different phases of market development. 

To interpret the \hl{lending} patterns, we take a closer look at the monthly OG bins at the network and individual bank level. (Note that this goes beyond the previously established network measures of the individual maturity layers, as these are now merged according to whether they form a statistically significant lending pattern.) The notable result here is that the bins may be characterised by a simple aspect of the most important banks' individual lending behaviour: At monthly time scales the important banks in ``short'' bins tend to both lend and borrow equal amounts of cash (indicative of financial intermediation), while the important banks in the ``long'' bins tend to either lend or borrow (indicative of financing). For the development phases, this suggest that patterns of financial intermediation are present at all phases of development, while patterns more in line with financing only appear at later phases.

Our coarse-grained layered SBM of the ILN thus uncovers a correlation between statistically significant lending patterns between groups of banks, the economic functions of important banks, and the maturity classes of the loans involved, showing that maturity information matters for the understanding of ILNs.

\section*{Results}

The adopted notation is in line with the one of~\cite{peixoto2015, Peixoto2018a}. An ILN at month $t$ is denoted as $\{G_l\}$, with $G_l$ the network for one of the eight maturity layers: $G_\text{<1d}$, $G_\text{2-7d}$, $G_\text{8-30d}$, $G_\text{31-90d}$, $G_\text{91-180d}$, $G_\text{0.5-1y}$, $G_\text{1-3y}$, $G_\text{>3y}$. As we treat each month independently we do not attach a time label to $\{G_l\}$. The $G_l$ are directed weighted multigraphs. In addition to the asymmetrical adjacency matrix $A^l_{ij} \in \mathbb{N}_0$ of layer $l$, the $G_l$ possesses the unordered edge covariates $x_{ijk}^l$ ($k \in [1, A_{ij}^l]$ for $A_{ij}^l > 0$). Thereby, each $k$th parallel edge between banks $i$ and $j$ represents a loan lent from bank $i$ to bank $j$ with a maturity in maturity class $l$ and of size $x_{ijk}^l$. The \emph{collapsed network} $G_c$ corresponds to complete maturity aggregation. Its adjacency matrix $A_{ij} = \sum_l A^l_{ij}$ and the edge covariates $x_{ijk}$ are constructed by flattening (i.e.~collapsing) the $x_{ijk}^l$ along the $k,l$ axes. We denote a level of granularity by the maturity bin set $\{\ell\}$ where $\ell$ specifies a set of merged layers. For example, the OG of month 58 (see the last month in Fig.~\ref{OLP}(a)) corresponds to $\{\ell\} = \{\{\mty{<1d}, \mty{2-7d}\}, \{\mty{8-30d}\}, \{\mty{31-90d}\}, \{\mty{91-180d}, \mty{0.5-1y}, \mty{1-3y}, \mty{>3y}\}\}$. The ILN $\{G_\ell\}$ representing the level of granularity specified by $\{\ell\}$ is constructed from $\{G_l\}$ by merging maturity layers according to $\{\ell\}$.

\subsection*{Descriptive statistics of the Russian ILN in a comparative framework}
We start by characterising the maturity layers and the collapsed form of the Russian ILN in terms of monthly (and occasionally yearly) time series of several ILN measures typically used in the literature. Layer analysis of layered ILNs has been performed for the interbank money markets of several countries, e.g.~ Mexico\cite{borboa2015} and the UK\cite{langfield2014}. The layers in those works, however, do \hl{(or can)} typically not differentiate between maturity classes\cite{Huser2015, Anand2018}. The maturity classes per layer of those that did are shown in Appendix G. \hl{The work by Bargigli et al.~\cite{bargigli2015}, which is related to this work, separates overnight loans, loans up to 1y and loans \textgreater 1y in a three layer end-of-year exposure network representation of the Italian ILN. They find that the structures of the three maturity layers are not representative of each other.} The measures that we analyse are: density, degree distribution, clustering coefficients, average shortest path length, degree mixing (as a proxy for bank size mixing), loan activity, and loan size. The results are compared to the stylised facts found in literature. A summary of this analysis is given in Table~\ref{stylized}, while details of the analysis can be found in the Supplementary Material, Appendix A.

\hl{The stylised facts describing prototypical ILNs are often deduced from exposure networks with only one maturity class.} Looking at the maturity layers separately, we find layer non-homogeneity for all measures in Table~\ref{stylized} except for the distribution of degree and transaction volumes. This means that variations of the ILN measures across the maturity layers are observed. While the stylised features are valid for the short maturity layers ($G_\text{<1d}$, $G_\text{2-7d}$, $G_\text{8-30d}$) and for the collapsed network ($G_c$), they become progressively invalid with growing loan maturities. As a matter of fact, we find that the stylised facts of the collapsed Russian ILN do not hold over all maturity layers. Given the similarities in the stylised facts across countries, we anticipate a similar behaviour for the ILNs of other countries. The short layers $G_\text{<1d}$, $G_\text{2-7d}$, $G_\text{8-30d}$ contain $97$\% of all issued loans. Upon merging those we more or less retrieve the collapsed Russian ILN with all the loan issuances. The longer maturity loans represent a few percent of the issuance but they are of sizeable economic relevance due to their specific turnover and their weight in the outstanding loans (see Fig.~\ref{phase-loans}(c)). During times of turmoil the short-term loans are often not renewed, but the long-term ones remain on the books till maturity, making them important for the stability of the ILN. \hl{This is reflected in the interest rate spreads shown in Appendix H.}

Core-periphery (CP) structure has been observed in many real-world networks~\cite{Borgatti1999} and in ILNs~\cite{Huser2015, Caccioli2018}. The seminal work of Craig and von Peter\cite{Craig2014} introduced an economic foundation for its occurrence in ILNs, i.e.~the elementary function of economic intermediation performed by banks. CP structure breaks with traditional theoretical banking literature where the interbank money market is modelled as a centralised exchange in which banks smooth out liquidity shocks. In contrast to the centralised exchange model, an ILN with CP structure gives rise to a sparse network. Thereby, a group of densely connected ``core'' banks perform the economic function of financial intermediation between numerous smaller, sparsely connected ``periphery'' banks. Formally, an ILN has CP structure if the lending patterns can be fully explained by grouping the banks into either ``core banks'' or ``periphery banks''. In the ideal situation, the bilateral relations between the banks define the \hl{bank} group memberships (i.e. whether a bank is core or periphery) by the following set of rules: (i) core banks lend to each other; (ii) periphery banks do not lend to each other; (iii) core banks lend to periphery banks; (iv) core banks borrow from periphery banks. For real-world ILNs with imperfect CP structure, several algorithms have been proposed (e.g.~\cite{Craig2014,dellarossa2013}) to detect CP structure and to determine the group memberships. We believe however that the proper way to establish CP structure in networks is by Bayesian inference of SBMs as these are ideally suited to parametrise the CP two-group structure and the four rules mentioned above, rather than minimising an objective function that might detect spurious CP structure, as shown by some recent literature~\cite{bargigli2015,lip2011fast,Kojaku2018}. Inferring CP structure by Bayesian inference of SBMs has been proposed in~\cite{Zhang2015} and applied to the Italian e-MID ILN in~\cite{barucca2016,Barucca2018} where depending on the time scale and SBM model extension, a bipartite or CP structure was found. We find some indications supporting a CP structure in the Russian ILN for $G_c$ and $G_\text{<1d}$, $G_\text{2-7d}$, $G_\text{8-30d}$. As explained in Appendix A these indications stem from the heavy-tailed degree distributions, disassortative degree mixing and the small average shortest path length.

\subsection*{Modelling the Russian ILN with the coarse-grained layered SBM}

The idea of banks behaving in groups with respect to lending and borrowing because of trading relationships in the interbank money market has been posited in various forms in the literature~\cite{krugman1996self,Furfine2001,Iori2008,veld2014,Craig2014,Craig2015,Huser2015,Braeuning2017,Blasques2018}, though not often explicitly in the form of SBMs\cite{barucca2016,Barucca2018,Caccioli2018}. Such group structure, also called network mesoscale structure, is abundant in real-life complex networks\cite{Newman2003}, notably social networks\cite{holland1983stochastic}. In the Russian interbank money market, there are several reasons to anticipate that the group structure is important: relationships are a way to solve problems with asymmetric information, a pervasive problem in the Russian banking sector and economy at large~\cite{Koen2013}; fragmentation of the Russian financial market due to the country's size (i.e.~eight time-zones); the presence of institutions controlled by the state to various degrees\cite{vernikov2016}. The most important advantages of using SBMs to detect the group or mesoscale structure are~\cite{peixoto2017bayesian}: (i) Theoretical guarantees against overfitting; (ii) They can be extended easily when formulated in a Bayesian setting; (iii) The ability to describe a wide variety of lending patterns (e.g. ILNs modelled by community structure, bipartite structure, CP structure or Erdős–Rényi graphs). As mentioned before, the SBM flavor we use to model the monthly ILNs is the coarse-grained layered SBM, which extends the layered SBM, itself an extension of the standard SBM. \hl{With ``coarse-grained layered SBM'' we refer to an extension to the layered SBM developed in~\cite{peixoto2015} which allows one to infer the OG along with the bank groups and other SBM parameters.} We shortly introduce \hl{the SBM, the layered SBM and the rationale behind its coarse-grained extension in a qualitative setting} before discussing the coarse-grained layered SBM \hl{for the Russian ILN}. A comprehensive discussion about SBMs in a Bayesian setting can be found in~\cite{peixoto2017bayesian} and \hl{the coarse-grained layered SBM used in this work is presented} in~\cite{peixoto2015,Peixoto2018a}.

\subsubsection*{\hl{Introduction to the SBM, the layered SBM and the coarse-grained layered SBM}}

SBMs are canonical models to study clustering and perform community detection~\cite{Abbe2017, young2018}. SBMs model topological patterns by assuming that the nodes ``behave group-like'' rather than on individual account, i.e. for a network \emph{$G$} with $N$ nodes modelled by $B \leq N$ groups, one assumes that the amount of connections between any two nodes $1 \leq i, j \leq N$ depends only on their group memberships $1 \leq b_i, b_j \leq B$, where $b_i$ is the group assignment of the $i$th node. When formulated in a Bayesian setting, the basic goal of SBMs is to determine the posterior probability distribution of all possible group assignments $\{b_i\}$ (where $B = \max_i b_i$) given the observed network $G$, a quantity written as $p(\{b_i\}|G)$. Because this is intractable for networks with more than a few nodes and edges, one is typically content with the maximum a posteriori probability (MAP) estimate\hl{, i.e.} $\argmax_{\{b_i\}} p(\{b_i\}|G)$, to which one refers to as ``the fit'' to the observed network $G$. Maximising the posterior $p(\{b_i\}|G)$ in search for the MAP estimate equivalently minimises the information-theoretic \emph{description length} (DL) of the data $G$, i.e. \hl{$\Sigma = -\log p(G, \{b_i\}) = \mathcal S + \mathcal L$ with $\mathcal S = -\log p(G|\{b_i\})$ and $\mathcal L = -\log p(\{b_i\})$. Choosing the base of the log to be 2,} $\mathcal S$ is the number of bits needed to describe the data ($G$) given the model parameters \hl{$\{b_i\}$} and $\mathcal L$ is the number of bits necessary to describe the model parameters. In other words, the best fit to the data is the one that \emph{compresses} it most, i.e.~yields the shortest DL. This is the minimum description length principle (MDL).

\hl{Though MDL as a regularisation device is fully equivalent to Bayesian model selection~\cite{mackay2005,peixoto2014hierarchical}, we invoke the MDL principle in our qualitative discussion} as it provides an arguably more intuitive explanation to how SBMs formulated in a Bayesian setting achieve robustness against overfitting \hl{when the model and the prior probabilities accurately represent our (lack of) knowledge~\cite{peixoto2017bayesian,Jaynes2003}}. In the case of SBMs the \hl{primary parameter} that controls the model's complexity is the number of groups $B$. Increasing $B$ improves the \hl{maximum} likelihood fit $p(G|\{b_i\})$ \hl{monotonically}, as new groups become available to account for any possibly insignificant deviation from the group's behaviour. More complicated models (larger $B$) are only preferred if there is sufficient evidence available in the data to compensate the extra degrees of freedom. This is achieved in the Bayesian formalism by specifying a prior $p(\{b_i\})$ and subsequent model selection based on \hl{integrated likelihoods and} statistical significance. From the MDL view this robustness against overfitting is achieved in the following manner: If $B$ becomes large, it decreases $\mathcal S$ but increases $\mathcal L$. The latter functions as a ``penalty'' that disfavours overly complex models~\cite{Peixoto2016}. The optimal choice of $B$ minimises the DL $\Sigma$, which induces a proper balance between $\mathcal S$ and $\mathcal L$. In other words, the optimal choice of $B$ and $\{b_i\}$ corresponds with the model that compresses the data most.

Layered SBMs are extensions of SBMs that additionally allow the group behaviour to depend on the network layers. In this work we use a specific layered SBM known as the independent layers SBM \hl{for all monthly ILNs}. \hl{The independent layers SBM assumes that a layered network $\{G_l\}$} can be modelled as one group structure which exhibits a topological pattern in each layer. In other words, each layer \hl{$G_l$} is modelled by an independent SBM constrained by the fact that the group memberships of the nodes \hl{$\{b_i\}$} are the same across all layers. \hl{Thus the model complexity of a layered SBM is now additionally controlled by the number of layers $L$ present in the observed layered network $\{G_l\}$ ($1 \leq l \leq L$), next to the number of groups $B$. The increased model power relative to the standard SBM again raises the question of overfitting: given the group structure $\{b_i\}$ of the layered network $\{G_l\}$, is it necessary to posit $L$ different topological patterns for each layer $G_l$, or can some layers be explained equally well by just one topological pattern and hence be merged? The layered SBM itself cannot provide an answer to this question, as $L$ is determined by $\{G_l\}$ and is thus simply a fixed component of the model complexity.}

\hl{The coarse-grained layered SBM extends the layered SBM by assuming that the observed ``high resolution'' layered network $\{G_l\}$ may be explained by an underlying ``lower resolution'' layered network $\{G_\ell\}$ consisting of the merged $G_l$ according to $\{\ell\}$, a set of layer bins which specifies the level of granularity of $\{G_\ell\}$. The optimal level of granularity (OG) is inferred simultaneously with the number of groups $B$ and the group structure $\{b_i\}$ by searching for the layered SBM that most compresses the \emph{lower resolution} layered network $\{G_\ell\}$, while taking into account the inevitable information loss incurred due to the lower degree of granularity (i.e. decrease in quality of fit). When the OG for a given layered network $\{G_l\}$ is complete aggregation, this indicates that the layer divisions in $\{G_l\}$ do not correlate with the mesoscale of its associated collapsed form $\{G_c\}$.} \hl{By contrast,} an OG that is different from complete aggregation points to a \hl{mesoscale structure} that is too complicated to be understood at the collapsed network level. The OG merges layers \hl{such that} the layer bins in $\{\ell\}$ ``acquire meaning'' so that at the bin interfaces the topological patterns between the groups change in a statistically significant way.


\subsubsection*{The coarse-grained layered SBM \hl{for the Russian ILN}}

We approach the Russian ILN as a time series of monthly ILNs and model each month separately with the \hl{coarse-grained layered SBM introduced in the previous section}. The specific flavor of the \hl{underlying} layered SBM that we use is the microcanonical independent layers weighted DCSBM, which takes the following features (above the expected group structure captured by the standard SBM) into account:
\begin{itemize}
\itemsep 0pt
\item
Edge directedness and the possibility of parallel edges, i.e.~multiple loans can be made between two banks in a given month.
\item
Heavy-tailed degree distributions (see Appendix A). This is captured by the degree-corrected SBM (DCSBM)~\cite{Karrer2011}.
\item
Heavy-tailed loan size distributions~\cite{Vandermarliere2015}. This is captured by extending the DCSBM to the weighted DCSBM\cite{Peixoto2018a} where the sizes of the loans between bank groups are modelled by log-normal distributions. In this way each ordered pair of bank groups has a lending pattern modelled as consisting of loans whose size's magnitude has a characteristic scale~\cite{Jaynes2003}.
\item
Maturity classes, modelled as network layers. This is captured by the independent layer SBM. The lending pattern in each layer is modelled by an independent weighted DCSBM with the constraint that the bank groups are identical in each layer.
\end{itemize}


\hl{Because of computational limitations, we do not explicitly take into account variables such as balance sheets, bank ownership, and interest rates. We deem, however, that this does not severely impact the ability of our model to capture the structural variability in the network. Indeed, structural features originating from the omitted variables can be effectively captured by the model. First, while balance sheets are not explicitly taken into account, we include degree distribution, which correlates to bank size, and loan size distribution (see Appendix A). Second, if bank ownership drives significantly different lending- and borrowing behaviour, these patterns can still be captured by the SBMs without explicitly including the information. It would be interesting to study if and how bank ownership correlates to the function banks fulfil in the different maturity layers. This falls outside the scope of this work. A study of the term structure of interest rates is included in Appendix H. In line with the expectation theory of interest rates~\cite{russell1992}, the yield-curve is upward sloping for longer maturities. Longer loan maturities lead to higher interest rates through averaging expected future short-term rates and adding a premium for liquidity- and default-risk. The introduction of maturity layers effectively captures the fact that differing risk patterns manifest themselves through differing loan structures in the layers. Recent work on the generative processes of risk in ILNs has shown that information such as interest rate and Credit Default Swap (CDS) spreads~\cite{ECB_CDS,Morrison2016}, play a central role in contagion dynamics. We note that securitised products, CDO, and CDS play a very marginal role in the Russian ILN in our period under study. In this way the Russian ILN provides a natural lab where the control parameter of risk derivatives can be effectively set to zero. In the forthcoming discussion on possible future research directions those opportunities will be highlighted. Local structures such as dense subgraphs fall beyond current SBMs' potential to capture mesoscale structure~\cite{peixoto2017bayesian}.} In the Methods section we motivate the choice to treat the \hl{maturity layers of the} monthly ILNs as independent.

The coarse-grained layered SBM augments the parameter set of the layered SBM $\{\theta\}$ by the maturity bin set specifying the level of granularity $\{\ell\}$. Thus its parameters are denoted as $(\{\theta\}, \{\ell\})$.
The coarse-grained layered SBM of a monthly ILN $\{G_l\}$ is a generative model given by~\cite{peixoto2015}
\begin{equation}
  p(\{G_l\}, \{\theta\}, \{\ell\}) = p(\{G_l\} | \{\theta\}, \{\ell\}) \times p(\{\theta\}) \times p(\{\ell\})\hl{.} \label{cglSBM} \; 
\end{equation}
Expressions for the model likelihood $p(\{G_l\} | \{\theta\}, \{\ell\}) \propto p(\{G_\ell\}|\{\theta\})$ and prior probabilities $p(\{\theta\})$ and $p(\{\ell\})$ can be found in~\cite{peixoto2015,Peixoto2018a} where $p(\{G_\ell\}|\{\theta\})$ and $p(\{\theta\})$ are defined by the layered SBM and $p(\{G_l\} | \{\theta\}, \{\ell\})$ and $p(\{\ell\})$ are defined by the coarse-grained extension. As the underlying maturity classes are inherently ordered, the \hl{uninformative} prior probability for the maturity bin set $p(\{\ell\})$ is determined by the constraint that only contiguous layers may be binned. In Appendix B we infer the OGs under the more general non-contiguous binning assumption (with a different prior $p(\{\ell\})$) and find qualitatively similar results as in Fig.~\ref{OLP}.

The posterior probability of the parameters $\{\theta\}, \{\ell\}$ is proportional to Eq.~\ref{cglSBM}:
\begin{equation}
  p(\{\theta\}, \{\ell\}|\{G_l\}) = \frac{p(\{G_l\}, \{\theta\}, \{\ell\})}{p(\{G_l\})}, \label{post}
\end{equation}
where $p(\{G_l\})$ is independent of $\{\theta\}$ and $\{\ell\}$. We may infer the bank groups $\{b_i\} \in \{\theta\}$ and the OG as the maximum a posteriori probability (MAP) estimate by searching for the mode of Eq.~\ref{post} (or equivalently Eq.~\ref{cglSBM}) with the inference algorithm explained in the Methods. In addition, we can compare the \hl{\emph{posterior odds ratio}} (POR) $\Lambda$ between \hl{two coarse-grained layered SBMs $\mathcal{M}_a, \mathcal{M}_b$ representing} two different levels of granularity $\{\ell\}_a, \{\ell\}_b$ by evaluating the ratio:
\begin{equation}
    \Lambda =  \frac{p(\{\theta\}_a, \{\ell\}_a, \mathcal{M}_a|\{G_l\})}{p(\{\theta\}_b, \{\ell\}_b, \mathcal{M}_b|\{G_l\})}
            = \frac{p(\{\theta\}_a, \{\ell\}_a|\{G_l\})}{p(\{\theta\}_b, \{\ell\}_b|\{G_l\})}
            = \frac{p(\{G_l\}, \{\theta\}_a, \{\ell\}_a)}{p(\{G_l\}, \{\theta\}_b, \{\ell\}_b)}, \label{POR}
\end{equation}
where the constant $p(\{G_l\})$ and the prior beliefs $p(\mathcal{M}_a), p(\mathcal{M}_b)$ for \hl{the coarse-grained layerered SBMs} $\mathcal{M}_a$ and $\mathcal{M}_b$ have cancelled out, as we had no prior preference with regard to the \hl{degree of} granularity (i.e.~$p(\mathcal{M}_a) = p(\mathcal{M}_b)$). Values of $\Lambda > 1$ indicate that according to the data, $\{\ell\}_a$ is preferred over $\{\ell\}_b$ with a degree of statistical significance given by the magnitude of $\Lambda$~\cite{Peixoto2018a}. \hl{The model selection implicit in the POR can be illustrated as follows. Given that $\Sigma_a$ ($\Sigma_b$) denotes the DL of $\{G_l\}$ according to model $\mathcal{M}_a$ ($\mathcal{M}_b$) the above equation implies that  $\log \Lambda = \Sigma_b - \Sigma_a$ and one recovers the MDL principle. Indeed, the preferred model is the one that achieves the most optimal compression of  the data. Accordingly, the POR is a model selection criterion that operates similarly to alternate information-based ``goodness-of-fit'' criteria such as BIC~\cite{schwarz1978estimating} and AIC~\cite{akaike1974new}.  The POR criterion, however, is ``exact'' for the coarse-grained layered SBMs at hand while BIC and AIC rely on specific assumptions about the asymptotic shape of the model likelihood which are known to be invalid for the SBM~\cite{yan2014model,peixoto2017bayesian}.}
 
We use $\log _{10} \Lambda$ to determine confidence levels for rejecting complete differentiation and complete aggregation as OG for the monthly ILNs. This is achieved by setting in Eq.~\ref{POR} $\{\ell\}_b$ to the OG inferred from the algorithm and $\{\ell\}_a$ to either complete differentiation or complete aggregation. Note that these confidence levels of rejection give rise to values $\log_{10} \Lambda \leq 0$. Large negative values of $\log_{10} \Lambda$ point to strong evidence for rejecting complete differentiation and/or complete aggregation.

\subsubsection*{Example for a small layered network}

As an illustration, we determine the OG for a small network of 50 nodes with three generic interactions (A, B and C) drawn in bundles (Fig.~\ref{wrap}(a)). Instead of specifying the individual interactions, one can describe the network in a more parsimonious way by specifying the ``wiring patterns'' between groups of nodes. The nodes are grouped in circles, squares and triangles, so as to encode a specific high-level description of the network. For example, interaction type A does not occur between two squares and between two triangles. Interaction A gives rise to circle-triangle and circle-square interactions. There are no circle-circle interactions of the C type, and circle-circle interactions of the types A and B are sparse. This is the kind of higher-level organisation into groups of nodes that SBMs can infer.

The results of the fits with the coarse-grained layered SBM for three levels of granularity are displayed in Fig.~\ref{wrap}(b).
Comparing the DL $\Sigma$ for the three fits, the preferred model is the OG $\{\{\text{A},\text{B}\},\text{C}\}$ with $\Sigma = 1687\ \text{bits} \approx 211\ \text{bytes}$. The network in Fig.~\ref{wrap}(a) can be saved to disk in graph-tool's\cite{peixoto-gt} native binary format as a file with a size of approximately 3,400 bytes, while the OG coarse-grained layered SBM can compress this (e.g. using arithmetic coding) down to $\Sigma \approx 211\ \text{bytes}$ (excluding the bytes needed for storage of practicalities\cite{NRC1992} such as the file header). We use the POR $\Lambda$ of Eq.~\ref{POR} to determine the confidence levels. The model $\{\text{A},\text{B},\text{C}\}$ that stands for complete differentiation is rejected with $\log_{10}\Lambda \approx -80$, indicating that it is an overly complicated model of the group structure. Indeed, the wiring patterns in layers A and B can be summarised neatly by merging them, since the inferred groups in $\{\text{A},\text{B},\text{C}\}$ and $\{\{\text{A},\text{B}\},\text{C}\}$ are identical. In contrast, merging layers B and C induces a change in group structure where the distinction between the squares and triangles is lost and these two groups are aggregated into one with many internal interactions. It is worth mentioning that the $\{\text{A},\{\text{B},\text{C}\}\}$ model is still a more appropriate description of the network than the $\{\text{A},\text{B},\text{C}\}$ one.

\subsection*{Fitting the coarse-grained layered SBM to the monthly ILNs}

For each monthly ILN $\{G_l\}$ we fit the coarse-grained layered SBM by the MAP estimate of its parameters $(\{\theta\}, \{\ell\})$. From this we obtain a time series of the OGs and the bank groups $\{b_i\}$.

\subsubsection*{The monthly OGs}

Fig.~\ref{OLP} shows the monthly OGs together with the PORs relative to complete differentiation and complete aggregation. As noted before, each OG consists of a set of bins numbered from short to long maturity by the OG bin index (OGB index). The most important result is that complete aggregation is always rejected as the appropriate level of granularity for $\{G_{\ell}\}$. The maturity layers are thus informative of the monthly ILN's network structure in the sense that including them in a layered SBM yields an improved description compared to a \hl{layered} SBM of the collapsed monthly ILN $\{G_c\}$, \hl{because the lending patterns between the bank groups depend significantly on the maturity classes of the loans.} In other words, the lending patterns in the monthly ILNs $\{G_l\}$ correlate with the maturity classes in a way that cannot be captured completely by just considering the loans alone, i.e. without maturity information. This is also indicated by the fact that complete differentiation is seen to yield substantially better fits to $\{G_l\}$ than complete aggregation.
The $\log_{10} \Lambda$ of Eq.~\ref{POR} that measures the degree of rejection, tends to increase until roughly month 50 (February 2004), after which the degree of rejection becomes weaker. This aligns with the timing rumours surfaced about a large scale government investigation into money laundering by banks. This eventually caused several bank licenses to be withdrawn, see Appendix F for a time line.

The eight maturity classes reflect the \hl{CBR}'s reporting standards. One may therefore ask whether all eight classes also have an economic function in the actual lending and borrowing between banks. Interestingly, the OGs are always different from complete differentiation, which means that the maturity classes \hl{as defined by the CBR} are partly redundant and that the lending patterns may be described more effectively by merging maturity layers into bins according to the OG. The OGBs are indicative of the fact that lending patterns between the bank groups can be combined in a more comprehensive form. For example, the short layers ($G_\text{<1d}$, $G_\text{2-7d}$, $G_\text{8-30d}$) which are characterised by the ILN stylised features in Table~\ref{stylized} are almost always merged together. We emphasise that this does not necessarily indicate pointwise similarity between these layers; rather the merging of the layers in the OGB induce a new lending pattern between the bank groups that is significantly different from the other OGBs.

Figure~\ref{OLP}(a) shows that the number of OGBs increases with time, which points to a developing interbank money market as more significantly differing lending patterns emerge between the bank groups. An additional argument is that the OGs differ more and more from the complete aggregation (see Fig.~\ref{OLP}(b)). Indeed, the monthly number of bins in the OG correlate with the known two phases in the Russian money market's development: early development (before month 35) and emerging maturity (from month 35 onward) (see Methods). This indicates that the market's emerging maturity phase is characterised by a more complex layered SBM with up to four statistically significantly lending patterns between the bank groups. This is in contrast with the early development phase, where mostly only two lending patterns are discerned (maturities up to 30 days and maturities longer than 30 days). In other words, the interbank market can be characterised during the early development phase by the existence of only two distinct lending strategies between the bank groups.

\subsubsection*{The bank groups}

In Fig.~\ref{groups-NMI}(a) we display the number of groups $B$ $(1 \leq b_i \leq B)$ inferred for each monthly ILN. As $B > 1$ across time, the monthly ILNs do contain group structure. Second, $B$ follows the same upward trend as in Figs.~\ref{phase-loans}(a),(b);~\ref{OLP}(b), indicating that the increase in the number of groups goes hand in hand with the development of the Russian ILN into a more mature phase as the months in our dataset pass.

\hl{We now gauge the time correlations between the inferred monthly group structure by looking at the similarity between the inferred bank groups $\{b_i\}$ in consecutive months.} Fig.~\ref{groups-NMI}(b) shows the normalised mutual information (NMI) between the \hl{$\{b_i\}$} in \hl{a} given month $t$ and the previous month $t-1$. In the Russian ILN, one discerns non-important banks. They display little activity over time and volume, and accordingly they are inclined to fluctuate between groups. To account for this, we condition the NMI on $q$, a measure to control which bank strengths~\cite{barrat2004architecture} are included. The strength $s^l_i$ of a bank $i$ is defined by the total amount it borrows and lends in a layer $l$ during a certain month:
\begin{equation}
    s^l_i = s^{l,\text{in}}_i + s^{l,\text{out}}_i = \sum_{j,k} x^l_{jik} + \sum_{j,k} x^l_{ijk}\;. \label{eq:strength}
\end{equation}
We also define the size of a maturity layer $S^l = \sum_{i<j,k} x^l_{ijk}$ as the total amount borrowed or lent during a certain month. The size of the monthly ILN is given by $S = \sum_l S^l$. To construct Fig.~\ref{groups-NMI}(b), we calculate for each month $t$ the relative strength of each active bank $s_i = \sum_l s^l_i/2S$. Then we create a list of banks by adding one bank at a time, \emph{starting out with the strongest bank and proceeding in order of decreasing bank strength}\hl{, until the cumulative relative strength reaches $q$, and we include only the banks on this list in the analysis.} In this way, we exclude banks that are only responsible for an insignificant amount of lending and borrowing in the network. We intersect the included banks with those at time $t-1$ and calculate the NMI from the two intersected bank group memberships for $q = 0.95, 0.99, 1$. Figure~\ref{groups-NMI}(b) shows that with decreasing $q$ the NMI grows, pointing to increasing similarity between the inferred bank groups at consecutive months. It is hard to draw conclusions from the NMI without a reference. By excluding the ``5\% least important banks by strength'' ($q=1 \longrightarrow q=0.95$) a considerable increase in correlations between the inferred bank groups in consecutive months is observed. This is indicative of the temporal stability of the inferred groups.
\hl{This temporal stability emerges from the analysis without explicitly imposing intertemporal correlations in the algorithm. The temporal stability of the inferred groups is indicative for the robustness of the findings and corroborates the role of relationship lending in interbank markets. Explicitly modeling intertemporal correlations by means of a dynamic layered SBM model~\cite{peixoto2015,peixoto2017modelling} turned out to be computationally prohibitive when applied to our data. Another advantage of the adopted ``static'' layered SBM methodology is that it can be applied to both time series of ILN data and ILN data covering a specific time period. In addition, treating the monthly data independently provided an additional robustness check as the anticipated intertemporal correlations are recovered by the algorithm. Inversely, our results also indicate that intertemporal correlations can be exploited to reduce the computational complexity of the modelling of ``current'' ILN data by using the inferred structure of the ILN in preceding months as a reasonable initial estimate for the underlying structure.}

\subsubsection*{Characterisation of the OGBs}

The OGBs can be interpreted as corresponding with statistically significant differing lending patterns between the bank groups. For consistency, we numbered the bins in the OG with an OG bin index (OGB index) from short to long maturity. Even though their actual content can vary considerably through time, we find that this numbering scheme reveals surprisingly consistent patterns.

At the \hl{\emph{network level}}, Fig.~\ref{v} displays the OGB sizes $S^\ell = \sum_{l\in\ell} S^l$ throughout time. We see that the order of magnitude of the OGBs corresponding with the shorter maturities stays consistent across time, even though the composition of the OGBs (i.e.~the maturity layers in the OGBs) changes considerably through time. This is especially so for OGB 2 (see Fig.~\ref{OLP}(a)): it contains the long maturity layers during the early development phase but not during the emerging maturity phase. It is interesting to note that while the sizes of the ``longer'' OGBs are much smaller than the short ones, these are not merged in the emerging maturity phase of the Russian ILN. This points to lending patterns that are sufficiently different from those in the short OGBs. (This still holds, except in one month, for the non-contiguous binning in Appendix B.)

Finally, we look at the \hl{\emph{bank level}}, i.e. the lending behaviour of individual banks. Again we use the bank strength to single out the ``important'' banks. In Fig.~\ref{main:instrength} we have shown the time-integrated distribution of $s^{\text{in},\ell}_i/s^\ell_i$ of the top 10\% most important banks of each month, separately for each OGB $\ell$. (Note that the conclusions hold for other cutoffs -- see Appendix E.) The instrength of a bank $i$ in maturity bin $\ell$ during a certain month is $s^{\ell,\text{in}}_i = \sum_{l\in\ell} \sum_{j,k} x^l_{jik}$, and the strength $s^\ell_i$ is defined analogously as in Eq.~\ref{eq:strength}. At monthly time scales the important banks in ``short'' bins tend to both lend and borrow equal amounts of cash, while the important banks in the ``long'' bins tend to either lend or borrow, \hl{i.e. act either as sources or sinks of liquidity}. Together with the indications for CP structure in the short layers, this suggests that the economic function of the important banks changes from financial intermediation (short bins) to financing (long bins). The apparent patterns in the short bins are indeed reminiscent of the functions in CP structures, while those in the long bins have, to the best of our knowledge, not often been included in the literature. Thus our coarse-grained layered SBM of the ILN uncovers a correlation between statistically significant lending patterns between the bank groups, the economic functions of the important banks, and the maturity classes of the loans involved.

\section*{Discussion}


In this paper, we investigate the importance of loan maturity information in interbank lending networks towards understanding its mesoscale structure\hl{, i.e.~the higher-level organisation of the banks into groups}. We do this to better understand the possible diversity in lending and borrowing patterns \hl{between bank groups} in ILNs and their accompanying economic functions.
\hl{We find that the representation of ILNs common in the literature, where either one maturity layer is studied or an aggregated view on the maturity layers is used, is unable to fully uncover information on the diversity in structures and underlying functions of interbank loans. Even after introducing multi-directedness, degree distribution, and loan size distribution, the economic functions carried out by banks are obscured by neglecting maturity information in the ILN representation.}

We employ the complete population of all Russian interbank lending contracts over a \hl{57} months period (January~2000~-~October~2004, \hl{except January 2003}) containing uniquely granular loan maturity \hl{classes as defined by the CBR}. Descriptive statistics on the \hl{associated} maturity layers uncover non-homogeneity in network measures along loan maturity layers, indicating a diversity in lending and borrowing patterns.

To investigate the optimal maturity granularity, we apply a coarse-grained layered stochastic block model~\cite{peixoto2015} to the data. \hl{The subsequent analysis confirms the suboptimality of complete maturity aggregation for our dataset, as it obscures the existence of four maturity layer bins that contain significantly distinct lending- and borrowing patterns with various underlying functional economic interpretations.} We find, for example, a consistent shorter term maturity layer \hl{bin} that behaves in line with the theory around tiered banking, with important banks intermediating liquidity. We also detect another layer \hl{bin} that aligns with long-term financing of bank activities, with the important banks acting \hl{either} as sources or sinks of liquidity. 

These findings immediately imply that the common practice of complete maturity collapse or focus on a single maturity layer, by choice or for data limitations, obscures important information about the functions banks perform in an ILN. This leads potentially to unrealistic ILN models and misguided policy conclusions about systemic stability, especially so in times of liquidity crunches when the short term layer of the ILN issuance network collapses.\hl{These insights also align with and add to recent findings that banks use the different loan maturities to manage their duration gap, providing a direct link from interbank markets to financial stability and the allocative efficiency of the economy at large~\cite{Bluhm2016}.}

\hl{The lending patterns between the bank groups depend not only on the loan maturities }but also on the phase of development of the interbank money market. The longest maturity loans, for example, only show structure related to interbank financing at later phases of market development. This makes economic sense since long-term loans entail greater counter-party risk and counter-party trust is only established over time through engagement in long term relationships.

\hl{Our analysis builds on the strand of interbank literature that leverages the network representation of interbank systems. Thereby, one often starts from the topology of exposure networks while largely disregarding loan maturity and other meta-data on the banks and their connections. Our work conclusively illustrates that after accounting for distinct loan issuance and degree- and loan size distributions, loan maturity is still informative of structure and economic functions. Alternate approaches have specifically focused on systemic risk and contagion and concluded that funding risk and CDS spreads play a central role in the mechanisms of risk propagation. It remains an open question how the structural information in maturity layers could complement these approaches. The trust crisis in Russia has indicated that the occurrence of  a non-fundamentals crisis is clearly reflected in both the interest rates and the mesoscale structure of the maturity layers. This connection offers opportunities for future research.}

All in all our results imply that theoretical and empirical research can neither adequately grasp the generative process of ILNs nor arrive at reliable policy conclusions from ILN modelling and simulation in the absence of appropriate granular maturity information. This underlines the importance for policy and regulatory bodies to collect maturity information on interbank loans if they desire to arrive at reliable insights into the health and systemic stability of the interbank lending market. It should also stimulate further theoretical and empirical research that incorporates loan maturity in modelling of the ILN generative process, its dynamics and its occasional transition to phases of instability or collapse.

\section*{Methods}

\subsection*{Dataset}

\subsubsection*{Overview}

\begin{figure}[!ht]
\centering
\includegraphics{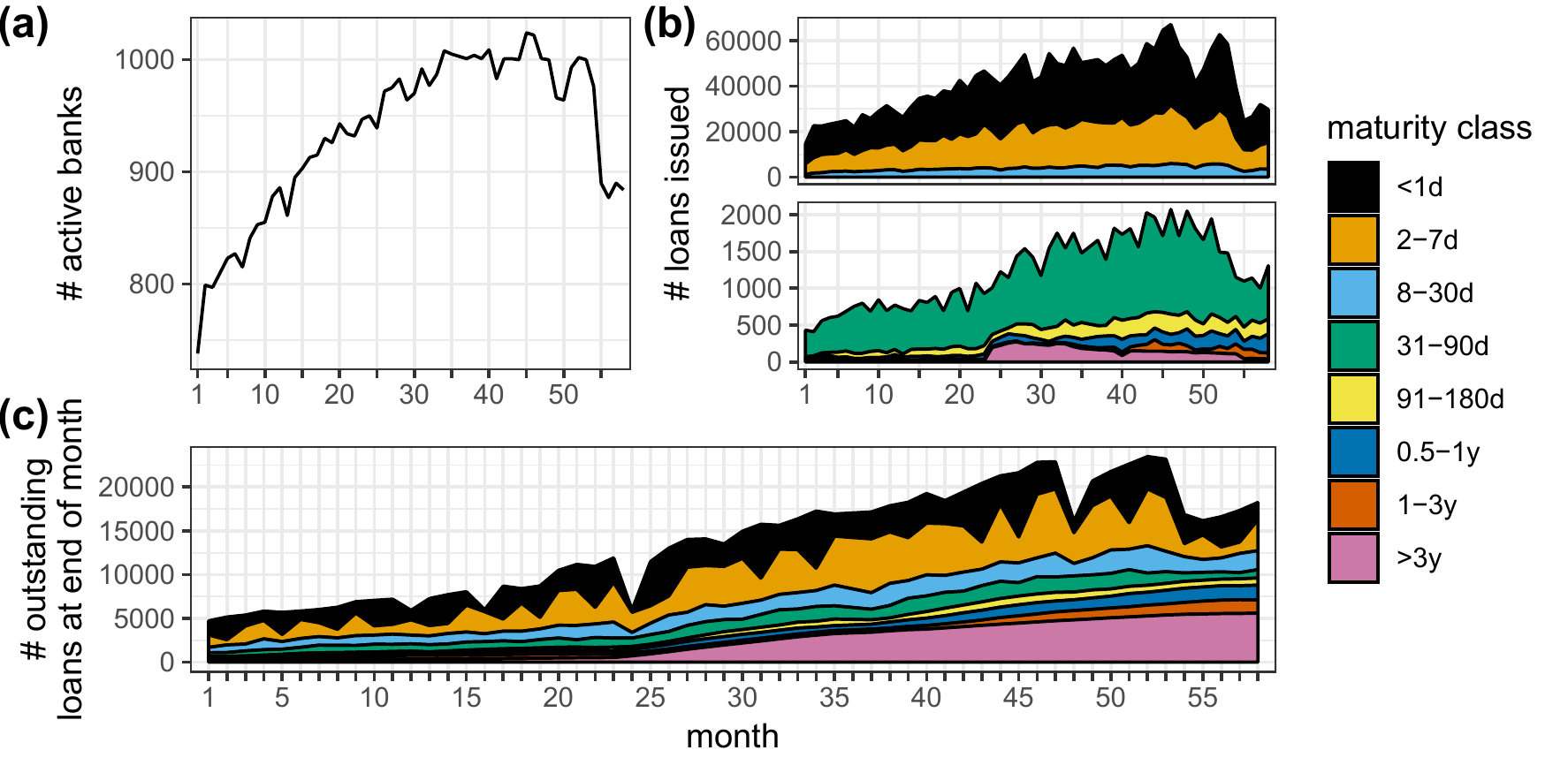}
\caption{Temporal evolution of the lending activity in the Russian interbank lending network. Month 1 corresponds with January 2000.
\textbf{(a)} The number of active banks on the lending market per month. A bank is active whenever it is the originator or the beneficiary of at least one new interbank loan in the given month. 
\textbf{(b)} The number of loans issued per month and per maturity. Note the different scales on the vertical axis of the two panels: The majority of issued loans have ``short'' ($\le$ 30 days) maturities. 
\textbf{(c)} The number of outstanding loans, i.e.~loans open on the last working day of each month. From this perspective the loans with longer maturities now play an equally important role as the short ones. Note that we have included loans issued before January 2000 for this panel (see Methods).\label{phase-loans}}
\end{figure}

\begin{figure}[!ht] 
\centering
\includegraphics{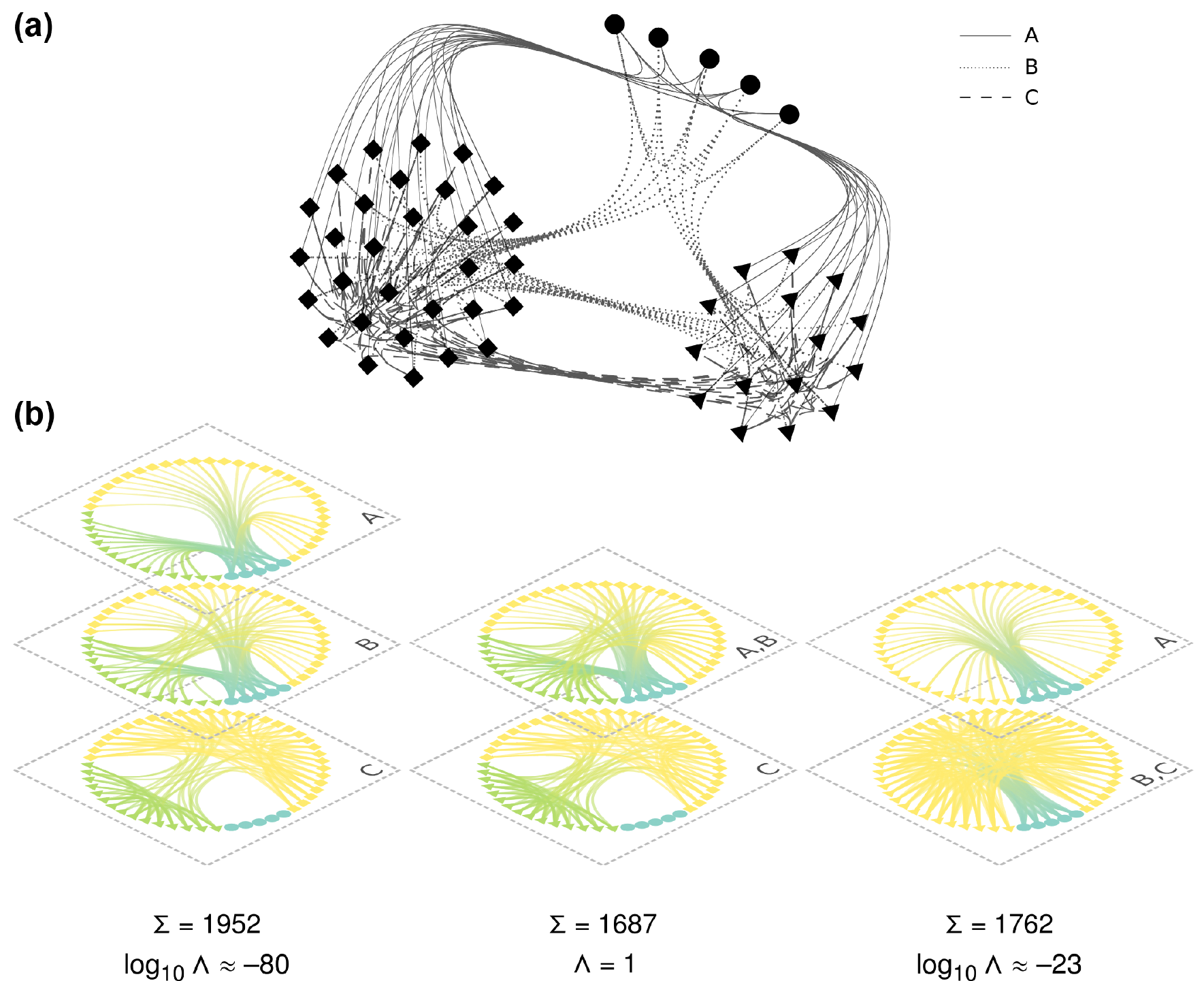}
\caption{Illustration of inferring the optimal granularity (OG) for a small network with the coarse-grained layered SBM. Images created with the graph-tool Python library\cite{peixoto-gt}. 
\textbf{(a)} An undirected and unweighted network with three types of edges (representing three generic interactions of type A, B, C) and three types of nodes (circle, square, and triangle). The interaction types define the network layers. Their network structure can be described as: perfect core-periphery (CP) (layer A), imperfect CP (B), and community structure (C).
\textbf{(b)} The network is shown with three levels of granularity. From left to right, these are: complete differentation $\{\text{A},\text{B},\text{C}\}$; merging of layers A and B with a lonestanding C $\{\{\text{A},\text{B}\},\text{C}\}$; and the merging of layers B and C with a lonestanding A $\{\text{A},\{\text{B},\text{C}\}\}$. The nodes are coloured according to the group index $b_i$ inferred by the coarse-grained layered SBM. The description length $\Sigma$ [bits] and posterior odds ratio $\Lambda$ relative to the OG for each representation are also indicated. The OG for this network is $\{\{\text{A},\text{B}\},\text{C}\}$. This can be understood by realising that both layers A and B have CP structure so that the merged layer $\{\text{A},\text{B}\}$ can be described more efficiently by just one CP model.
\label{wrap}}
\end{figure}

\begin{figure}[!ht]
\centering
\includegraphics{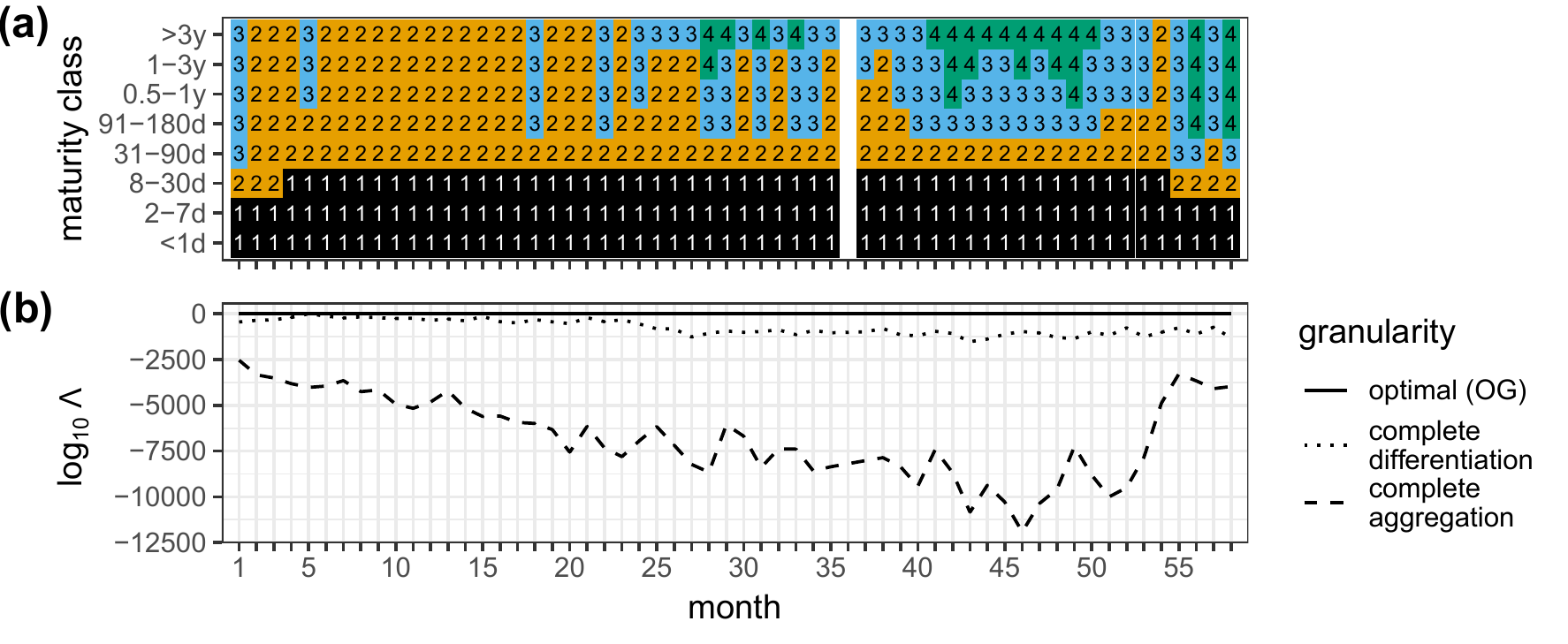}
\caption{The optimal granularity (OG) with respect to the loan maturity classes for the Russian interbank lending network. The OG corresponds to the maturity bin set $\{\ell\}$ parameter of the best-fit coarse-grained layered SBM. 
\textbf{(a)} Monthly time series of the OG inferred from the monthly interbank lending network. Each OG bin (OGB) holds one or more maturity classes and is labelled by an OGB index $1,2,3,4$ and indicated by a colour. The OGB index runs from short to long maturities. The OGBs correspond to lending patterns between the bank groups that differ from each other in a statistically significant way.
\textbf{(b)} Temporal evolution of the $\log_{10}$ of the posterior odds ratio $\Lambda$ (for its definition see Eq.~\ref{POR}) of the \hl{coarse-grained} layered SBM for three different granularities: (i) the OG; (ii) complete loan maturity differentiation; (iii) complete loan maturity aggregation. Complete aggregation is decisively rejected as an optimal representation of the \hl{bank group structure} of the monthly interbank lending network. The largest $\log_{10} \Lambda$-value different from zero is $\log_{10} \Lambda = -6.33$ and occurs for complete differentiation at $t=5$.\label{OLP}}
\end{figure}

\begin{figure}[!ht]
\centering
\includegraphics{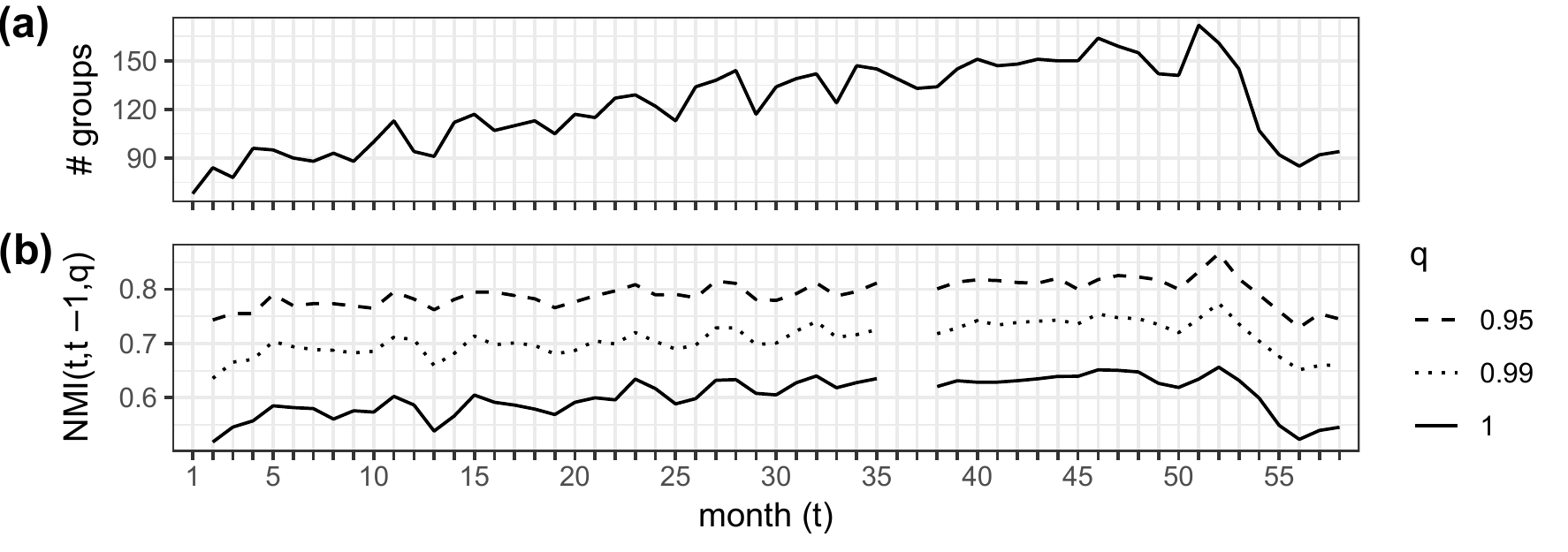}
\caption{Information about the bank groups $\{b_i\}$ for the monthly Russian interbank lending networks as inferred from the optimal granularity (OG). 
\textbf{(a)} Temporal evolution of the number of different bank groups. The number of groups roughly increases until $t = 50$. This indicates that the mesoscale structure of the monthly interbank lending networks becomes progressively more complex as more bank groups are needed to fit the lending patterns. 
\textbf{(b)} The temporal evolution of the normalised mutual information (NMI) between the OG group memberships of the set of banks active in two consecutive months $(t, t-1)$ for three strength fractions $q$. For example, for $q = 0.95$, banks responsible for $95$\% of the network's lending and borrowing are included. In many fields, including community detection, the NMI is a popular quantity to measure the similarity between two partitions of a set~\cite{Fortunato2010}.\label{groups-NMI}}
\end{figure}

\begin{figure}[!ht]
\includegraphics{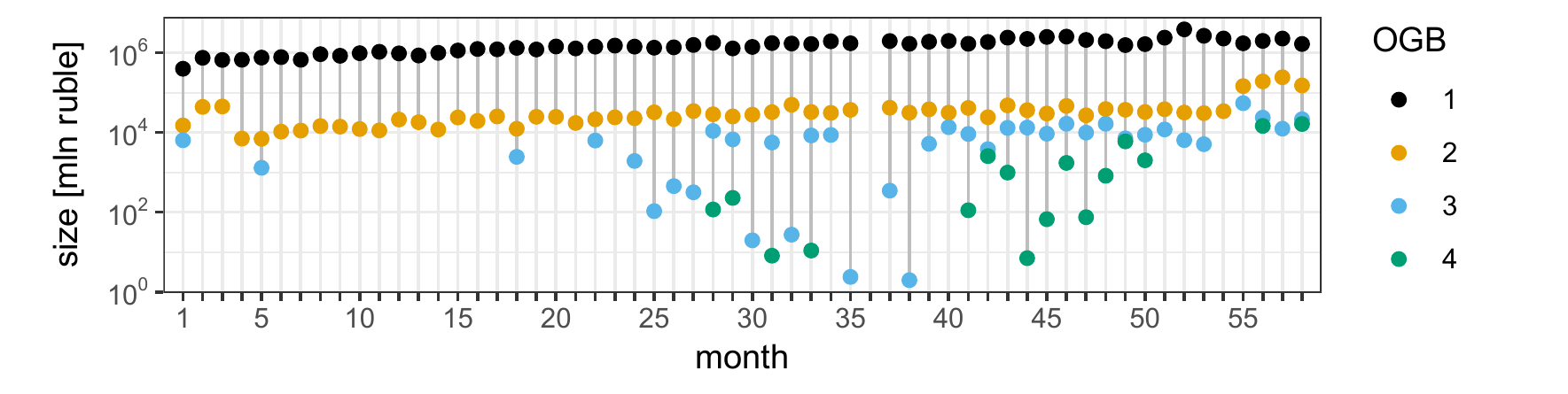}
\caption{Correlations between aggregate loan sizes and lending patterns: the size of the optimal granularity bins (OGBs) through time. The size of an OGB is defined as the total amount of issued money in the interbank lending network defined by the OGB. The OGB indices and colours correspond to those of Fig.~\ref{OLP}(a).
\label{v}}
\end{figure}

\begin{figure}[!ht]
\includegraphics{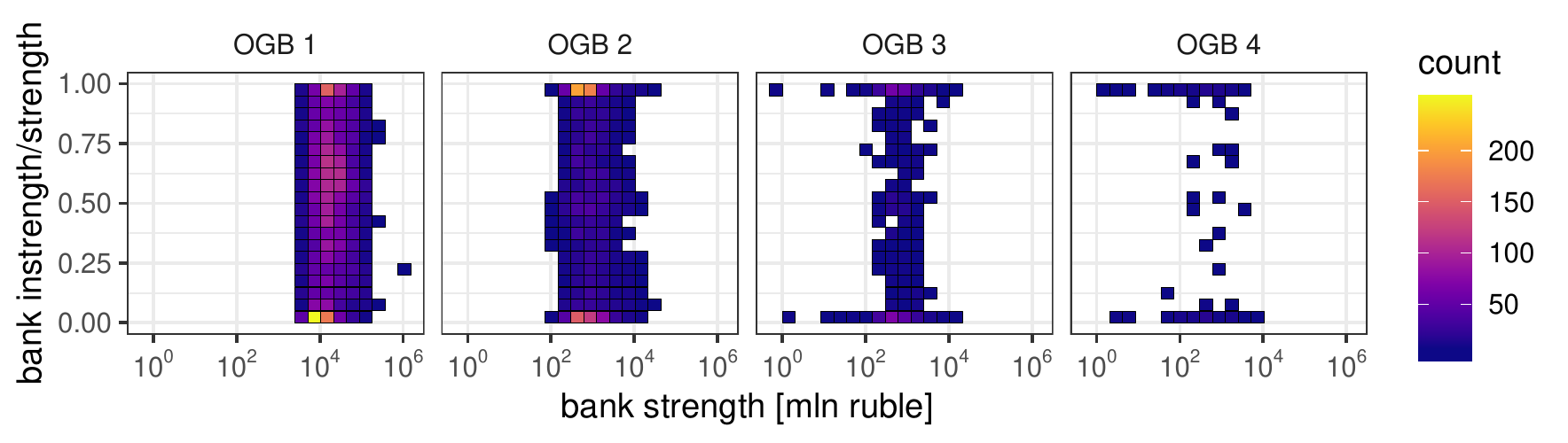}
\caption{The time-integrated distribution of the strength and the instrength/strength ratio of the monthly ``important banks'' for the four optimal granularity bins (OGBs). As in Fig.~\ref{groups-NMI}(b) we gauge a bank's importance by its strength: For a given month a bank is deemed important if its strength lies in the top 10\%. With growing OGB index, the important banks increasingly tend to either lend or borrow. Banks in OGB 1 and OGB 2 tend to balance lending and borrowing. This suggests that the economic function of the important banks in the various OGBs changes from financial intermediation (short maturity bins) to financing (long maturity bins) on a monthly time scale.\label{main:instrength}}
\end{figure}

The interbank data analy\hl{s}ed in this work offers a rich and unique account of Russian commercial bank activities over a six-year timeline (August 1998 - October 2004). The data is provided by Schoors and Karas~\cite{Karas2010} and has been painstakingly assembled from public and private sources. \hl{It originates directly from the information reported to the CBR about interbank contracts.} The dataset used in this paper can be retrieved on demand from the authors. To the best of our knowledge the dataset is unique~\cite{Anand2018} in quite a number of aspects. We mention the availability of information about the issuance months of loans (most often only the derived exposure is available) and rather detailed granular information about the maturity (see Appendix G). The fact that the Russian interbank market went through many stages of development during 1998-2004 adds an additional layer of dynamics. Note that the CBR is not included as a node and that we do not have information about any of its transactions. A classification of the banks together with a recent description of the money market can be found in~\cite{vernikov2016}. On average, about half of the Russian banks are active on the interbank market~\cite{Vandermarliere2015}. The dataset starts a few weeks before the so-called ``1998 Russian default''~\cite{Chiodo2002}, which caused a complete collapse of the interbank money market. In response the CBR imposed to little avail several extraordinary measures to stabilise the market. These exceptional circumstances greatly disrupted the workings of the interbank market and because of this we have restricted our analysis to the period of 57 months from January 2000 until October 2004. \hl{In our numbering scheme month 1 corresponds to January 2000, month 13 to January 2001, and month 58 to October 2004.} As can be inferred from Fig.~\ref{phase-loans}, during months 1-35 the market is developing. We refer to this period as the early development phase. Roughly starting from month 35 (December 2003) the interbank market enters an emergent maturity phase: the amount of active banks stabilises and the share of overnight loans declines in favour of longer-maturity loans. The emergent maturity phase includes the trust crises in the second half of 2003 and the summer of 2004 caused by a money laundering scandal (see Appendix F for more details). As compared to the Russian default of 1998, the crises of 2003 and 2004 were less disruptive for the interbank lending market.

\subsubsection*{Contents}

The data consists of the issuance of domestic unsecured interbank loans, annotated by lender bank ID, borrower bank ID, month of issuance, loan size, interest rates and maturity class. \hl{As noted earlier, the loan interest rates are not incorporated into the coarse-grained layered SBM; we discuss them in Appendix H.} The loan issuances are reported to the legislator on the first day of each month throughout August 1998 up until November 2004 in one of the following eight maturity classes: overnight (less than one day, \mty{<1d}), 2-7 days (\mty{2-7d}), 8-30 days (\mty{8-30d}), 31-90 days (\mty{31-90d}), 91-180 days (\mty{91-180d}), 0.5-1 year (\mty{0.5-1y}), 1-3 years (\mty{1-3y}), more than 3 years (\mty{>3y}). Because of this reporting standard, the precise issuance month of the loans is known. For January 2003 (month 36) there is no data available. Whenever possible, we interpolate the missing month in the time series. For some variables (Figs.~\ref{OLP}(a) and \ref{groups-NMI}(b)) this is not feasible and this is at the origin of the missing data point in the time series.

\subsubsection*{Aggregation window}

Loans are aggregated by issuance month into monthly ILNs (as in~\cite{Vandermarliere2015}) for two reasons. First, monthly aggregation is the most granular time scale available in the data. Second, monthly compliance with regulatory requirements for banks (e.g.~for liquidity and capital) induces a monthly periodicity in the data.

\subsubsection*{Availability}

The dataset as well as the inferred bank groups are available on request.

\subsection*{Implementation}

\subsubsection*{Inference algorithm with agglomerative hierarchical clustering}

We infer the OG for a layered network using an agglomerative hierarchical clustering heuristic as suggested in~\cite{peixoto2015}. At the start of the procedure each layer is put in its own bin. In the next steps, bins are merged so as to reduce the overall DL. The overall DL consists of the DL of the layered SBM plus an ``SBM extension term'' that accounts for the model selection between the possible levels of granularity (Eq.~17 in~\cite{peixoto2015}). With contiguous binning only contiguous bins are merged, while for non-contiguous binning any pair of bins may be merged at each step. In this way a series of layered networks is generated, starting with the original layered network and ending with the collapsed network. The layered network with the smallest overall DL defines the OG.

We use the efficient inference algorithm~\cite{peixoto2014} implemented in the graph-tool library~\cite{peixoto-gt} to fit the layered SBMs to the monthly ILNs. The algorithm employs an agglomerative heuristic to fit the layered SBM and a multilevel Markov Chain Monte Carlo to sample the posterior distribution. We used High Performance Computing resources to perform four runs of numerical calculations: Two of those used contiguous binning and two of those used non-contiguous binning for the maturity layers. 
In each run we used agglomerative hierarchical clustering to fit the layered SBM and to sample the posterior for the levels of granularity generated for each of the monthly ILNs. The number of samples were set to 10,000 and 25,000 for both the contiguous and non-contiguous runs. All runs (including the test runs) yielded similar results for each monthly ILN which alludes to the stability of the posteriors. The layered SBMs with smallest DLs were gathered from both runs.

\section*{Acknowledgements}
We would like to thank Benjamin Vandermarliere for his invaluable input and comments while writing the paper, and Bart de Boer for supporting this collaboration. This research was supported by the Research Foundation Flanders (FWO) under Grant Number G018115N and G015617N, and Bijzonder Onderzoeksfonds BOF2452014000402.

The computational resources (Stevin Supercomputer Infrastructure) and services used in this work were provided by the VSC (Flemish Supercomputer Center), funded by Ghent University, FWO and the Flemish Government -- department EWI.

\section*{Author contributions statement}

Author contributions: M.V.S., M.v.d.H, J.R, and K.S. designed research; M.V.S., and M.v.d.H performed research; M.V.S. analysed and processed results; and M.V.S., M.v.d.H, J.R, and K.S. wrote the paper.

\section*{Additional information}

\textbf{Competing interests:} \hl{The authors declare no competing interests.}

\newpage
\renewcommand{\refname}{References}

\newpage

\begin{table}[ht!]
\caption{Stylized network properties of interbank lending networks (ILNs) according to selected studies. For some ILN measures conflicting values are reported, in which case we separate them by a backslash. For example, low clustering coefficients are reported by~\cite{Finger2013}. The values for the ILN measures in bold apply to the collapsed Russian ILN (see~\cite{Vandermarliere2015} and Appendix A). With the layer homogeneity we indicate whether the quoted ILN measures apply to all maturity layers of the Russian ILN.\label{stylized}}
\centering
\begin{tabular}{@{}llll@{}}
\toprule
ILN measure   & Value & Layer homogeneity  & Selected studies \\ \midrule
density                      & \textbf{sparse}  & No        & \cite{Finger2013,Huser2015,aldasoro2016,Blasques2018,silva2015,bargigli2015}        \\
degree distribution (in and out degrees)          & \textbf{heavy-tailed} & Yes   &  \cite{Finger2013,Huser2015,aldasoro2016,Vandermarliere2015,bargigli2015,Craig2014}       \\
topological structure                     & scale-free / core-periphery &  &\cite{demasi2006,soramaki2006,bargigli2015}/\cite{Huser2015,Craig2014,aldasoro2016,Blasques2018,silva2015} \\
clustering coefficients      & low / \textbf{high} & No   &  \cite{Finger2013}/\cite{Huser2015,aldasoro2016,soramaki2006,bargigli2015} \\
average shortest path length & \textbf{small} / "small world" & No   &  \cite{aldasoro2016,silva2015}/ \cite{Finger2013,Huser2015,soramaki2006}       \\
bank size mixing             & \textbf{disassortative} & No & \cite{Finger2013,Huser2015,aldasoro2016,demasi2006,silva2015,borboa2015}       \\
distribution of transaction volumes           & \textbf{heavy-tailed}   & Yes & \cite{Finger2013,Vandermarliere2015}       \\ \bottomrule
\end{tabular}
\end{table}

\clearpage

\newpage
\goodbreak
\thispagestyle{empty}
\section*{\Huge{Appendix}}
\tableofcontents
\appendix

\clearpage

\section{Maturity layer analysis}

Here we give a detailed description and network analysis of the Russian interbank lending network to compare to other interbank networks used in research. We divide the Russian interbank loan network $\{G_\alpha\}$ by maturity class $\alpha$ into separate layers consisting only of the loans with loan maturity inside each maturity class. (Note that in the main text we use the symbol $l$ instead of $\alpha$.) We evaluate the properties per maturity layer (often also referred to as layer for simplicity) to show the differences between them. We denote the eight maturity layers as: $G_{<1d}$, $G_{2-7d}$, $G_{8-30d}$, $G_{31-90d}$, $G_{91-180d}$, $G_{0.5-1y}$, $G_{1-3y}$, $G_{>3y}$.

\subsection*{Activity and volume}
One expects that the number of banks participating in a given loan maturity layer decreases as the loan maturity lengthens. Table~\ref{numbanks} shows that this is approximately correct. More importantly, one sees that the number of active banks becomes very small for $G_{1-3y}$ and $G_{>3y}$. We define an active bank as a bank that has lent or borrowed at least once in a given time window and maturity layer. We will always only consider active banks to deal with the death and birth of banks in the Russian interbank market over the course of time~\cite{Karas2005}. For the layers with long loan maturity, this results in an exceedingly high average lending activity per bank, comparing to layers with shorter loan maturity. It seems that $G_{>3y}$ is occupied by a small club of banks that trade relatively intense.

\begin{table}[ht]
\centering
\caption{(top panel) The number of active banks involved in loans by maturity layer and year. A total of 1,040 banks are present in the data. (lower panel) The number of loans per active bank by maturity layer and year, obtained by dividing the top panel of Table~\ref{numloans} and the top panel of this Table element-wise. Care must be taken when interpreting; this is not indicative of an `average' bank since the degree distributions are heavy-tailed (as will be discussed below). Also note that the columns `1998' and `2004' are biased because they are incomplete in the data (respectively 5 and 10 months missing months). \label{numbanks}}
\begin{tabular}{@{}lrrrrrrr@{}}
\toprule
\multicolumn{1}{l}{} & \multicolumn{1}{c}{1998} & \multicolumn{1}{c}{1999} & \multicolumn{1}{c}{2000} & \multicolumn{1}{c}{2001} & \multicolumn{1}{c}{2002} & \multicolumn{1}{c}{2003} & \multicolumn{1}{c}{2004} \\ \midrule
\textless1d& 610& 810& 878& 954& 1,022& 1,027& 983\\
2-7d& 589& 907& 953& 1,026& 1,074& 1,098& 1,058\\
8-30d& 375& 849& 920& 980& 1,018& 1,054& 986\\
31-90d& 166& 535& 648& 698& 706& 775& 686\\
91-180d& 63& 203& 265& 310& 321& 429& 378\\
0.5-1y& 49& 120& 132& 169& 204& 258& 259\\
1-3y& 24& 64& 80& 60& 58& 113& 107\\
\textgreater3y& 11& 45& 29& 38& 33& 23& 20\\ \bottomrule
\end{tabular}

\vspace{.25cm}

\begin{tabular}{@{}lrrrrrrr@{}}
\toprule
\multicolumn{1}{l}{} & \multicolumn{1}{c}{1998} & \multicolumn{1}{c}{1999} & \multicolumn{1}{c}{2000} & \multicolumn{1}{c}{2001} & \multicolumn{1}{c}{2002} & \multicolumn{1}{c}{2003} & \multicolumn{1}{c}{2004} \\ \midrule
\textless1d          & 13                       & 118                      & 187                      & 258                      & 286                      & 337                      & 225                      \\
2-7d                 & 7                        & 57                       & 108                      & 161                      & 183                      & 229                      & 147                      \\
8-30d                & 3                        & 15                       & 32                       & 43                       & 44                       & 59                       & 43                       \\
31-90d               & 1                        & 5                        & 10                       & 11                       & 15                       & 18                       & 12                       \\
91-180d              & 1                        & 2                        & 3                        & 4                        & 5                        & 7                        & 5                        \\
0.5-1y               & 1                        & 2                        & 2                        & 4                        & 4                        & 7                        & 7                        \\
1-3y                 & 2                        & 3                        & 3                        & 3                        & 3                        & 8                        & 8                        \\
\textgreater3y       & 5                        & 3                        & 6                        & 10                       & 78                       & 74                       & 44                       \\ \bottomrule
\end{tabular}
\end{table}

Table~\ref{numloans} lists the yearly, and average monthly lending activities by loan maturity class, where we define the \emph{lending activity} simply as the number of loans recorded during a certain period. A crucial observation is that the lending activity sharply decreases for longer maturities, which is also reported for other interbank markets in~\cite{bargigli2015}. We see that the overnight segment ($G_{<1d}$) is the most active, together with $G_{2-7d}$. The impacts of the crises in 1998 and 2004 have had clear impact on the lending activity, except for the longer maturities; these seem relatively unaffected.

\begin{table}[ht]
\centering
\caption{(top panel) Lending activity, defined as number of of loans, by loan maturity class and year. Note that 1998 and 2004 are incomplete years in the data, counting 5 and 10 months respectively. (lower panel) Monthly average lending activity by loan maturity layer. Time series of the lending activity can be found in Fig.~1 in the main text.\label{numloans}}
\begin{tabular}{@{}lrrrrrrr@{}}
\toprule
\multicolumn{1}{l}{} & \multicolumn{1}{c}{1998} & \multicolumn{1}{c}{1999} & \multicolumn{1}{c}{2000} & \multicolumn{1}{c}{2001} & \multicolumn{1}{c}{2002} & \multicolumn{1}{c}{2003} & \multicolumn{1}{c}{2004} \\ \midrule
\textless1d & 7,917 & 95,389 & 163,855 & 245,840 & 292,048 & 346,221 & 220,979\\
2-7d& 4,325& 51,673& 103,376& 165,602& 196,765& 251,006& 155,834\\
8-30d& 958& 12,914& 29,250& 42,067& 45,204& 62,190 & 42,181 \\
31-90d& 245& 2,770& 6,365& 7,939& 10,606& 14,184& 8,443\\
91-180d& 82& 460& 742& 1,189& 1,754& 2,797& 2,010\\
0.5-1y& 63& 237& 307& 593& 886& 1,901& 1,899\\
1-3y& 52& 207& 279& 202& 180& 914& 845\\
\textgreater3y& 55& 128& 180& 367& 2,588& 1,712& 890\\ \bottomrule
\end{tabular}

\vspace{.25cm}

\begin{tabular}{@{}lrrrrrrr@{}}
\toprule
\multicolumn{1}{l}{} & \multicolumn{1}{c}{1998} & \multicolumn{1}{c}{1999} & \multicolumn{1}{c}{2000} & \multicolumn{1}{c}{2001} & \multicolumn{1}{c}{2002} & \multicolumn{1}{c}{2003} & \multicolumn{1}{c}{2004} \\ \midrule
\textless1d& 1,583& 7,949& 13,655& 20,487& 26,550& 28,852& 22,098\\
2-7d& 865& 4,306& 8,615& 13,800& 17,888& 20,917& 15,583\\
8-30d& 192& 1,076& 2,438& 3,506& 4,109& 5,182& 4,218\\
31-90d& 49& 231& 530& 662& 964& 1,182& 844\\
91-180d& 16& 38& 62& 99& 159& 233& 201\\
0.5-1y& 13& 20& 26& 49& 81& 158& 190\\
1-3y& 10& 17& 23& 17& 16& 76& 84\\
\textgreater3y& 11& 12& 15& 31& 235& 143& 89\\ \bottomrule
\end{tabular}
\end{table}

Table~\ref{term-volume} lists the total loan volumes (i.e. loan sizes) by loan maturity layer and year. We observe that the relative importance of each maturity segment, as measured by the total volume of loans traded within it, follows the ranking of the loan maturity lengths. The loan volumes are log-normally distributed~\cite{Vandermarliere2015}, especially for shorter loan maturities.

\begin{table}[ht]
\centering
\caption{Total loan volumes by loan maturity and per year in billions of rubles. \label{term-volume}}
\begin{tabular}{@{}lrrrrrrr@{}}
\toprule
\multicolumn{1}{l}{} & \multicolumn{1}{c}{1998} & \multicolumn{1}{c}{1999} & \multicolumn{1}{c}{2000} & \multicolumn{1}{c}{2001} & \multicolumn{1}{c}{2002} & \multicolumn{1}{c}{2003} & \multicolumn{1}{c}{2004} \\ \midrule
\textless1d& 226.3& 2,674.4& 5,984.3& 8,855.9& 10,306.4& 13,844.6& 13,040.1\\
2-7d& 83.1& 1,030.4& 2,910& 5,114.7& 6,104.6& 9,058.1& 7,880.2\\
8-30d& 15.9& 220.7& 620.8& 1,103.1& 1,102.9& 1,871.8& 1,833.3\\
31-90d& 4.4& 31.9& 105.2& 205.3& 318.3& 415.4& 253.6\\
91-180d& 1.7& 5.3& 9.6& 23.4& 46.9& 73.6& 73.7\\
0.5-1y& 3.2& 2.4& 8& 20.2& 24& 50.8& 69.9\\
1-3y& 2.2& 5.5& 6& 6& 3.3& 10.1& 11\\
\textgreater3y& 2.6& 0.8& 1.2& 1.4& 1.3& 0.8& 2.5                      \\ \bottomrule
\end{tabular}
\end{table}

\clearpage
\subsection*{Maturity layer activity of banks}

\begin{figure}[ht]
\includegraphics{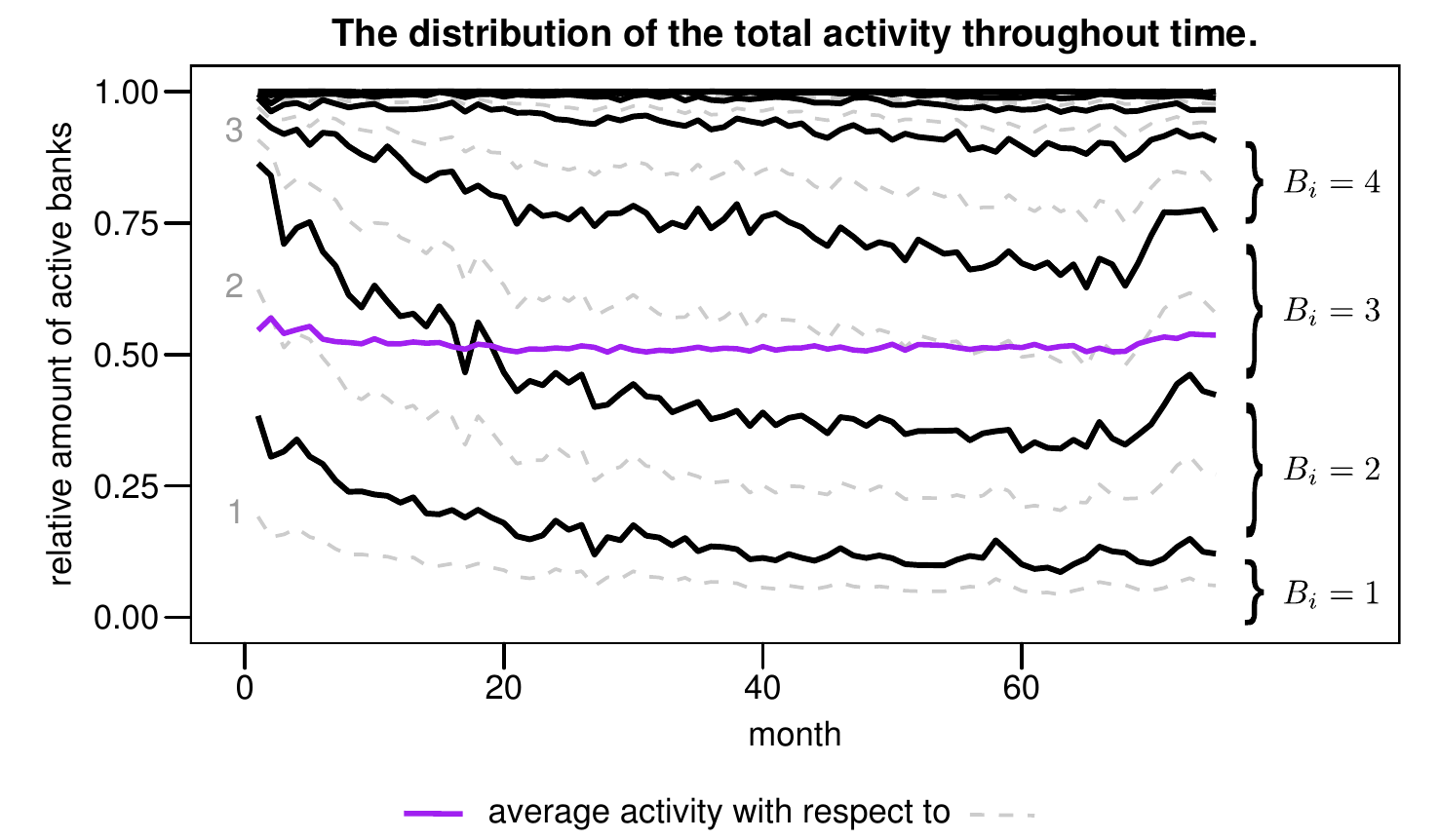}
\caption{For a given month, the area between the thick lines is proportional to the relative number of active banks with a certain $B_i$ value, which is indicated on the right. The dashed lines make up the moving coordinate system of $\langle B_{i} \rangle$, in relation to which the average total activity indicated by the thick purple line must be understood. For example, $\langle B_i \rangle$ is about 2 (3) for month 1 (75). The summer of 2004 crisis is mirrored by the drop in $\langle B_i \rangle$ during the last months. \label{multiplex-act}}
\end{figure}

A node $i$ is defined as active on layer $\alpha$ if it has at least one connection on this layer, i.e. its degree $k_i^\alpha > 0$. In symbols,
\begin{equation}
b_i^\alpha = \begin{cases}
   1 & \text{if } k_i^\alpha > 0,    \\
    0              & \text{otherwise.}
\end{cases}
\end{equation}
The \emph{total activity} $B_i$ measures the number of layers the node participates in, i.e. $B_i = \sum_\alpha b_i^\alpha$~\cite{Battiston2016}. Figure~\ref{multiplex-act} shows the distribution of $B_i$ on monthly time scales. We see that the average total activity grows steadily as the network develops, finally settling around at a value of about three. Almost no banks make use of more than six maturity layers on a monthly basis. The distribution of the total activity is quite broad and relatively unpeaked, especially during the stable phase of the network. This has been reported as a typical quality of real-world multiplex network~ \cite{Battiston2016a}.

\clearpage
\subsection*{Density}
Real-world interbank networks are typically \emph{sparse}, meaning that in directed and undirected views of the network only a small fraction of all possible edges exist. One can then ask how the edges are distributed among the banks. To this end, we define the \emph{degree} of a node. For undirected networks, this is the number of edges connected with a given node, i.e. the number of the bank's counterparties. For directed networks, the in-degree (out-degree) of a node is the number of incoming (outgoing) edges. The in-degree (out-degree) of a bank is then simply the number of loans borrowed (lent) by it.

\begin{figure}[ht] 
\includegraphics{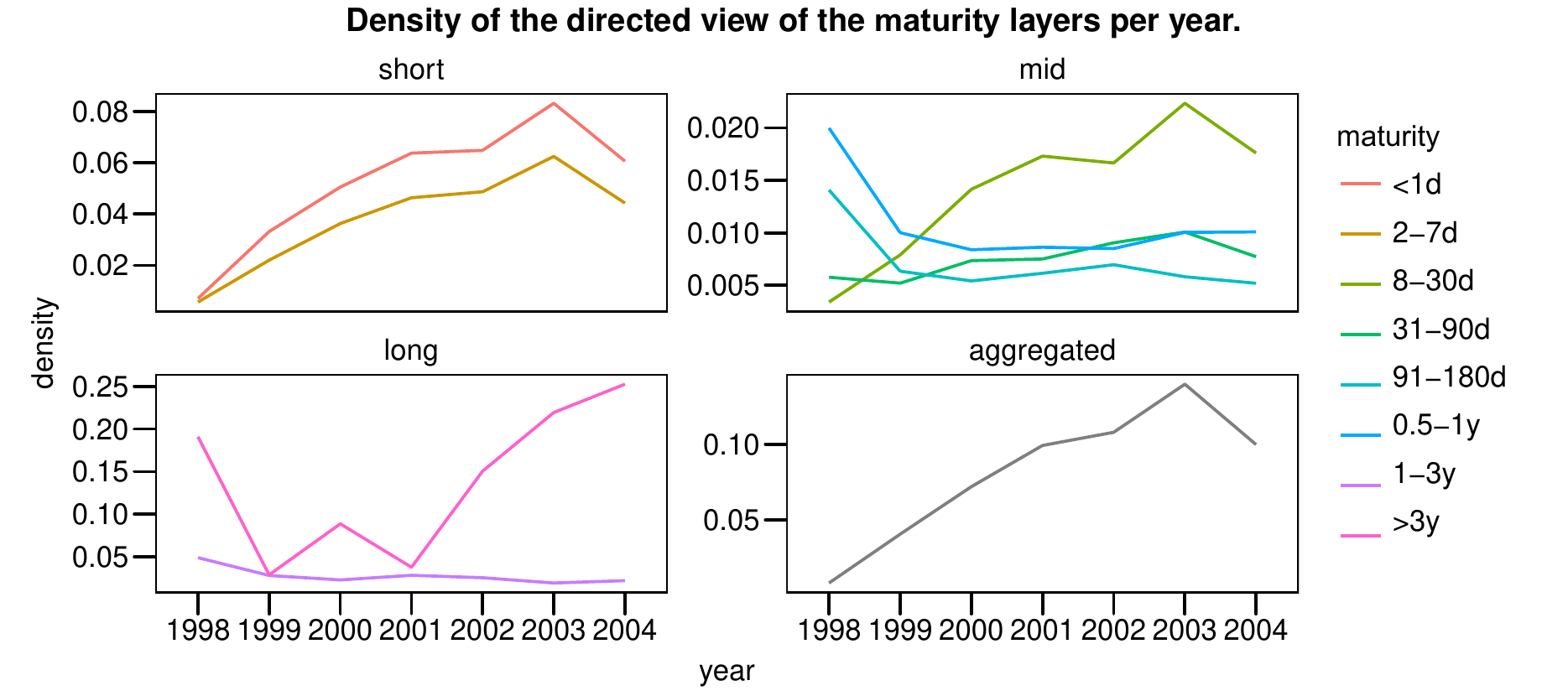}
\caption{The yearly density per loan maturity layer in the directed Russian interbank lending network. The directed density for a simple graph with $N$ nodes and $E$ edges is defined as $d = \frac{E}{N(N-1)}$. \label{clustering-dirdensity}}
\end{figure}

The above seen differences in lending and bank activity per loan maturity layer is reflected in the directed density of the layers, shown in Fig.~\ref{clustering-dirdensity}. All layers can be considered sparse, except for $G_{1-3y}$ and $G_{>3y}$, which have respectively moderate and an extremely high density~\cite{zlatic2009}. We also observe that the density in almost all layers grows steadily as the network develops, pointing to an increasing interconnectedness with increasing development of the market~\cite{VandenHeuvel2015}.

\subsection*{Degree distribution}

Interbank networks, like many real-world networks~\cite{newman2003}, exhibit heavy-tailed degree distributions. In a nutshell, this means that \emph{few nodes have many links and many nodes have few links}. In general, many typical distributions are heavy-tailed -- in fact, they abound in descriptions of natural events like avalanches, earthquakes, turbulent flow and rainfall~\cite{christensen2005,newman2005power}; another example in economics are the non-Gaussian return distributions in financial markets.

Early work in the previous decade tended to postulate power laws for the degree distributions. Networks whose degree distributions follow a power law (at least asymptotically) are called \emph{scale-free} networks. However, while agreeing on the heavy-tail character, recent literature has cast serious doubt on the idea that power laws are the best candidate for the degree distributions, and thus on the scale-free character of interbank networks. In our case, power laws have been decisively rejected as best fit candidates for the heavy-tailed degree distributions of the <1d and 2-7d loan maturity by Vandermarliere et al. (2015)~\cite{Vandermarliere2015}.

We plot the heavy-tailed degree distributions with complementary cumulative distribution functions (ccdfs)~\cite{newman2005power} with doubly logarithmic scales. If we denote the degree and its distribution by $k$ and $p(k)$ respectively, the ccdf is given by
\begin{equation}
\operatorname{ccdf}(k) \equiv 1 - \operatorname{cdf}(k) = P\left({k \leq k'}\right) = \int_k^\infty p(k') \ensuremath{\operatorname{d}\!{k'}},
\end{equation}
where $\operatorname{cdf}(k)$ is the usual cumulative distribution function.

\begin{figure}[h!] 
\includegraphics{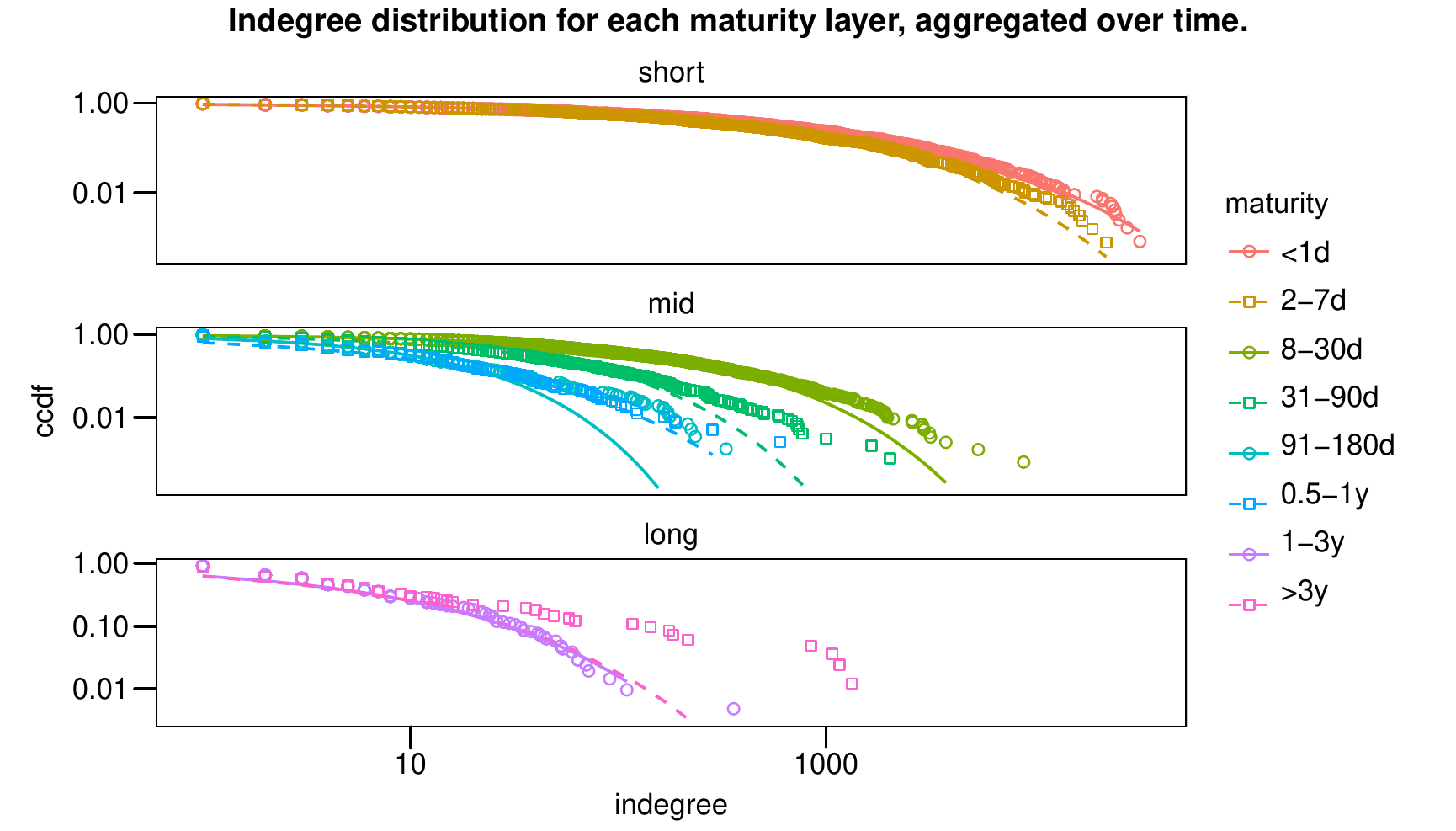}
\caption{In-degree distribution with attempts to fit the bulk+tail with stretched exponentials, drawn as piecewise linear functions connecting each predicted value. The correspondence with this distribution seems to break down for layers with loan maturity longer than 8-30d. Inferred $(\lambda, \beta)$ parameters for $G_{<1d}$, $G_{2-7d}$, $G_{8-30d}$ are $(2.6e-3, 4.3e-1)$, $(3.4e-3, 4.8e-1)$, $(1.0e-2,5.8e-1)$ respectively, with all standard errors below 1\%. \label{degrees-in}}
\end{figure}

\begin{figure}[h!] 
\includegraphics{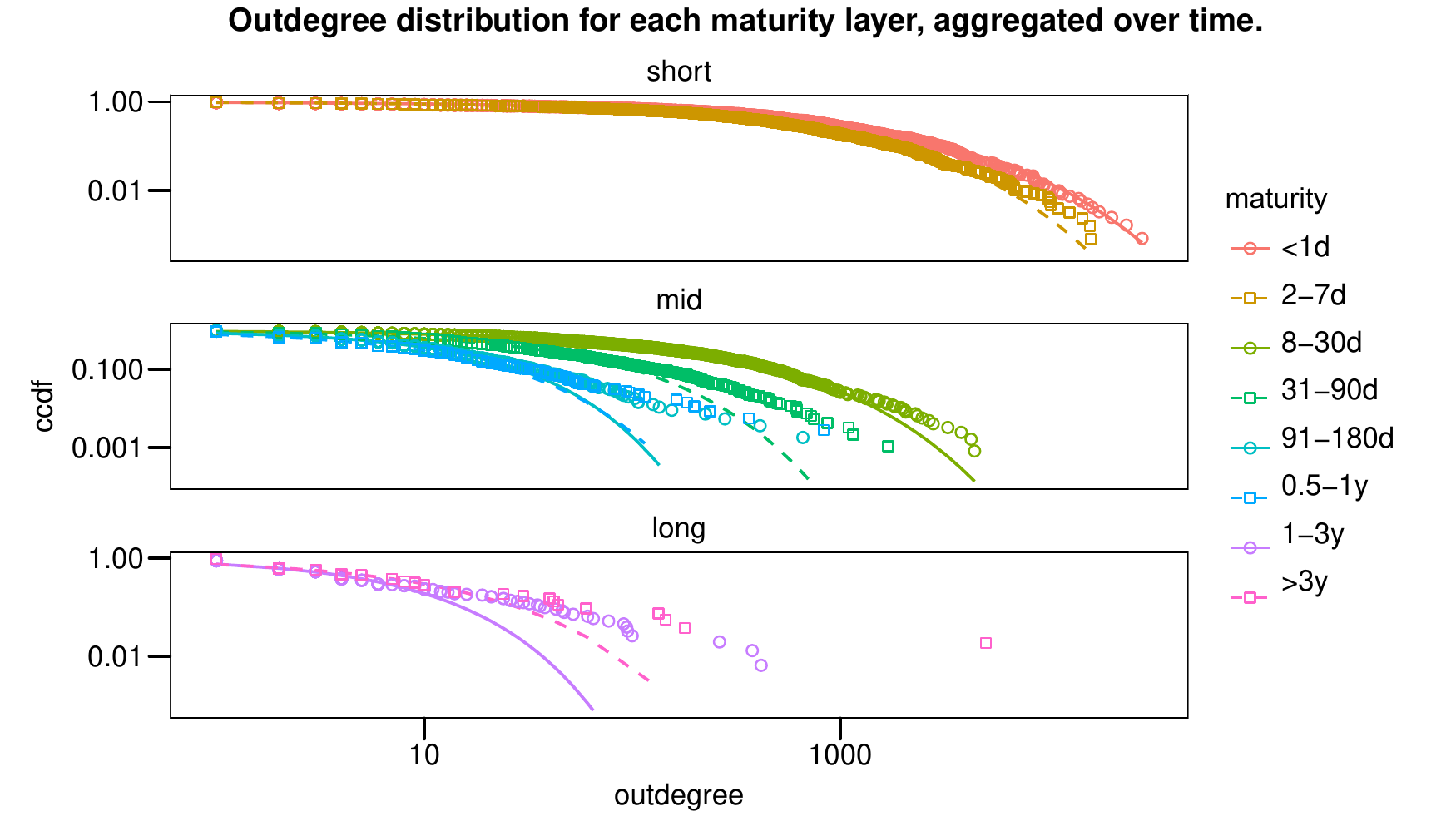}
\caption{Out-degree distribution with with attempts to fit the bulk+tail with stretched exponentials, drawn as piecewise linear functions connecting each predicted value. It is clear the correspondence with this distribution breaks down for layers with loan maturity longer than 8-30d. Inferred $(\lambda, \beta)$ parameters for $G_{<1d}$, $G_{2-7d}$, $G_{8-30d}$ are $(1.5e-3, 5.3e-1)$, $(2.3e-3, 5.7e-1)$, $(8.3e-3,6.1e-1)$ respectively, with all standard errors below 1\%. \label{degrees-out}}
\end{figure}

The degree distributions are shown per loan maturity layer in Figures~\ref{degrees-in} and~\ref{degrees-out}. Each loan maturity layer exhibits fat-tailed in- and out-degree distributions, with the maximum degrees separated some 5 ($G_{>3y}$, in-degree) to 18 ($G_{8-30d}$, in-degree) standard deviations from the mean. Vandermarliere et al. (2015)~\cite{Vandermarliere2015} established that stretched exponentials of the form
\begin{equation}
f(d) = C d^{\beta - 1} \exp{(-\lambda d)^\beta} \label{sexp}
\end{equation}
provide the best overall fit for the bulk+tail multi-directed in- and out-degree for monthly and yearly time windows. In Eq.~\ref{sexp}, $d$ is the in-degree or out-degree, $C$ a normalization constant and $\lambda, \beta$ distribution parameters. The stretched exponential can be understood as a Weibull distribution, with $\beta$ being the shape parameter, and $1/\lambda$ the scale parameter. If $0 < \beta < 1$, the distribution has a fat tail, with a smaller $\beta$ putting more weight towards smaller $d$. $1/\lambda$ widens the distribution, the mean of $f$ being proportional to it.

We fitted $f$ to the degree distributions to see if the result by Vandermarliere et al. (2015)~\cite{Vandermarliere2015} can be extended to longer maturities. The conclusions are identical for the in- and out-degrees. First, the stretched exponential fit does seem to describe the time-aggregated degree distributions of $G_{<1d}$, $G_{2-7d}$ and $G_{8-30d}$. Then the leap in residual sum-of-squares suggests that $G_{31-90d}$ and longer maturities can not be satisfactorily described by the stretched exponential. This is confirmed by visual inspection. While (parts of) the bulk distributions of the longer loan maturities are reasonably well captured by $f$, the fits systematically underestimate the ccdf in the tail; put differently, they underestimate for a given large degree $d$ how many nodes exists with an even larger degree $d'$.

\begin{figure}[ht]
\hspace*{-1cm}
\includegraphics{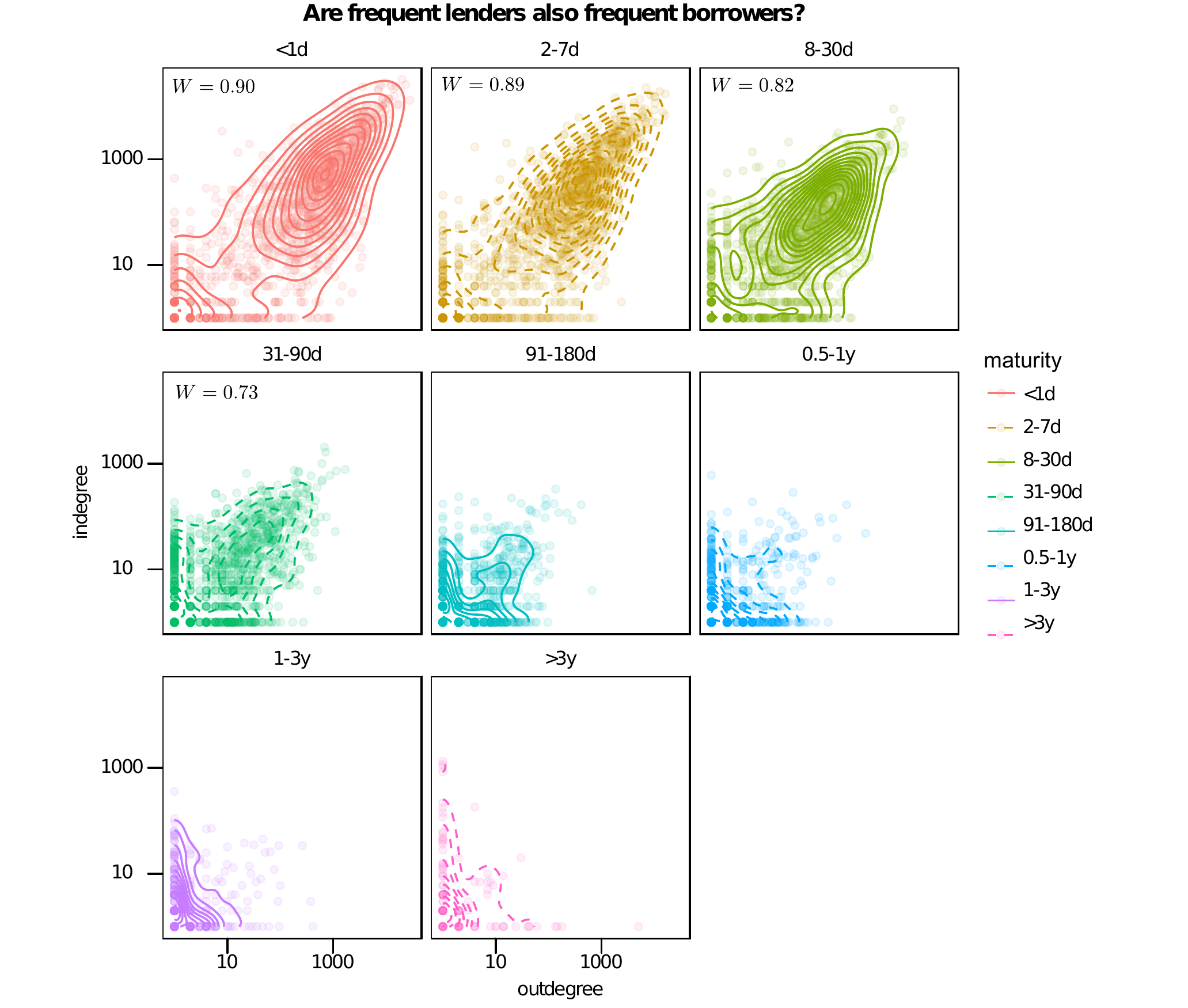}
\caption{The time-aggregated joint degree distribution for each layer, drawn as density contour plots with almost transparent data points. For the first four maturity bins, a highly significant Kendall $W$ was obtained and put in the top left corner of the panel. Note that all scales are equal for easy comparison. \label{degrees-inout}}
\end{figure}

To investigate whether frequent lenders are also frequent borrowers, we first rank the banks with regard to in- and out-degree, then we calculate Kendall's coefficient of concordance $W$, which measures the rate of agreement $0<W<1$ between two rankings, per loan maturity layer. The significant results together with the joint degree distributions are shown in Fig.~\ref{degrees-inout}. We conclude that lending and borrowing are highly correlated for $G_{<1d}$, $G_{2-7d}$, and $G_{8-30d}$; in other words, the dominant lenders are likely to be dominant borrowers, and vice versa. This correspondence breaks down markedly for the layers with longer loan maturity, with orderly decreasing $W$ but insignificant test results. This can also be seen in the shapes of the contour plots in Fig.~\ref{degrees-inout}: high in-degree - out-degree correlation is marked by squeezed contours around the plot diagonal, which does not hold for maturity longer than 31-90d. $G_{1-3y}$ and $G_{>3y}$ even display modest anti-correlation for the degree tails, meaning that the most frequent lenders (borrowers) area unlikely to be among the most frequent borrowers (lenders).

\clearpage
\subsection*{Average shortest path length}

The average shortest path length (denoted from here on as just `average path length') $D$ is defined for the largest connected component of the undirected view of interbank networks. It indicates the typical distance between two randomly chosen nodes, i.e. the smallest number of edges needed to reach one node from the other. \emph{Small-world networks} are characterized by high clustering and small average path length. Both should be compared to their mean values in an ensemble of random networks with the same number of nodes and edge~ \cite{watts1998}.

Most studies find that $D$ is small for interbank networks, which indicates compact network structure, but not all conclude that they are small-worlds. Hubs tend to lower the average path length; scale-free networks are ultra-small-worlds. $D$ is a measure of the typical length of intermediation chains that are taking place among the market participants (at least in the largest connected component). Longer intermediation chains arise when $D$ is large, which effectively contribute to slowing down the market transactions between participants and consequently harming the liquidity allocation between financial institutions. In contrast, when $D$ is small, the information between the market participants flows quickly in the network, giving rise to a well-functioning liquidity allocation in the market~\cite{silva2015}.

In the case of the Russian interbank network, the average path length peaks at times when the network is in crisis, and decreases gradually when the network is maturing. The same patterns have been reported in other empirical literature, such as in~\cite{Blasques2018}.

\begin{figure}[h!]
\includegraphics{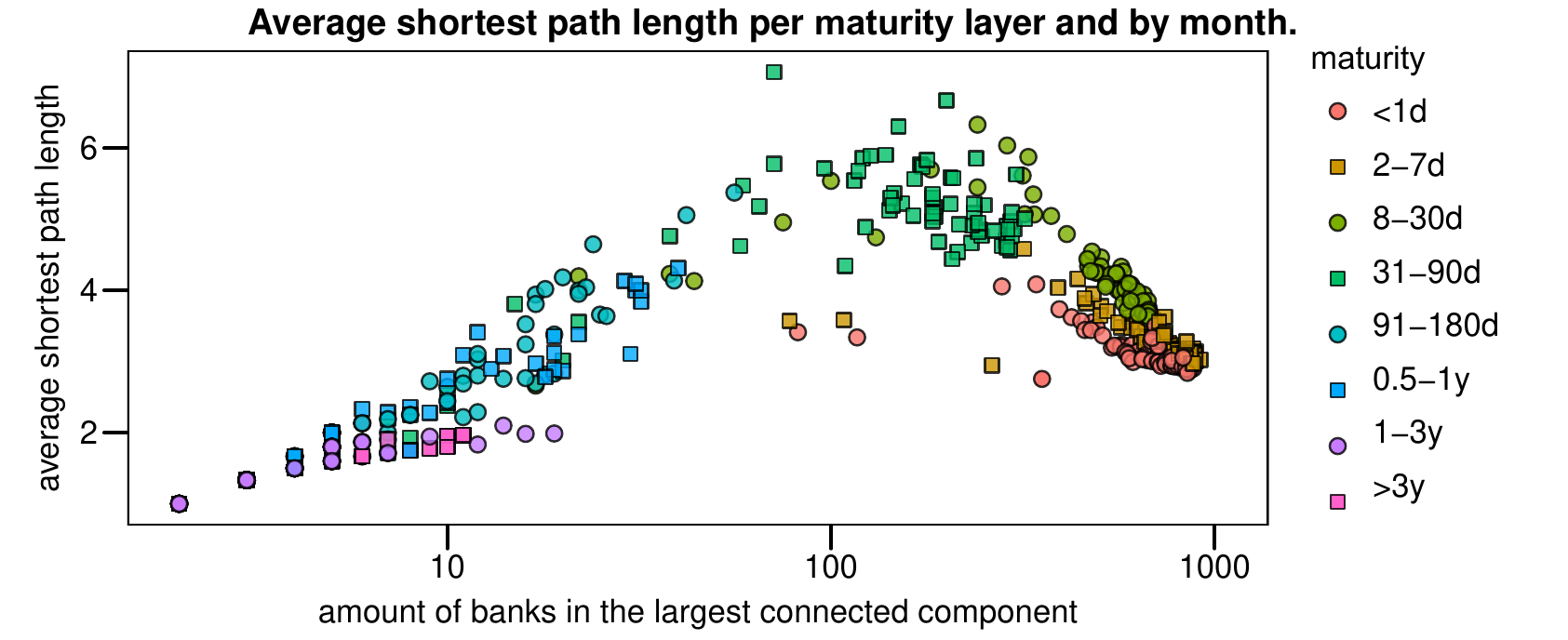}
\caption{The average (shortest) path in function of the size of the largest (weakly) connected component. For the four shortest loan maturity layers, the points lying astray from the dense cloud all occur during the first five months. Note that the number of banks in the largest connected component for these layers is notably larger than those of the remaining layers. \label{distance-month}}
\end{figure}

\begin{figure}[h!]
\includegraphics{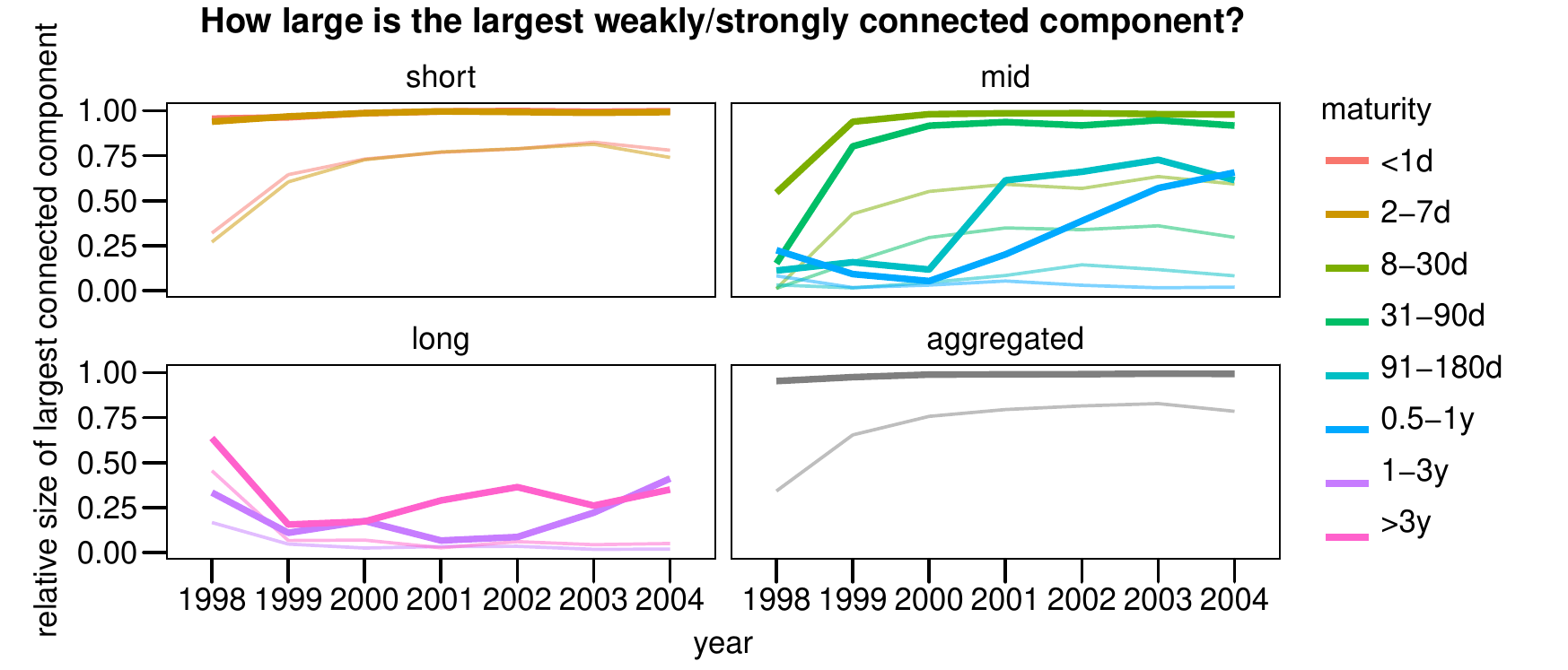}
\caption{The relative size of the largest connected component, given the weak or strong requirement, is calculated by dividing the number of banks in the largest connected component by the total number of active banks in a specific layer and year. The fat (thin) lines indicate the weak (strong) version. \label{clustering-gcsize}}
\end{figure}

Figure~\ref{distance-month} shows that exclusively for the layers with loan maturity <90d the average path length $D$ \emph{decreases} roughly linearly when the size of the largest connected component (LCC) increases \emph{logarithmically}. Herein the LCC covers the vast majority of the active banks within yearly time windows, see Fig.~\ref{clustering-gcsize}. Thus we see that core-periphery structure, which we expect in these layers from our analysis in the Topology section, is extremely effective in channeling and intermediating liquidity; of course this comes at the cost of systemic risk in the core hubs.

\begin{figure}[h!]
\includegraphics{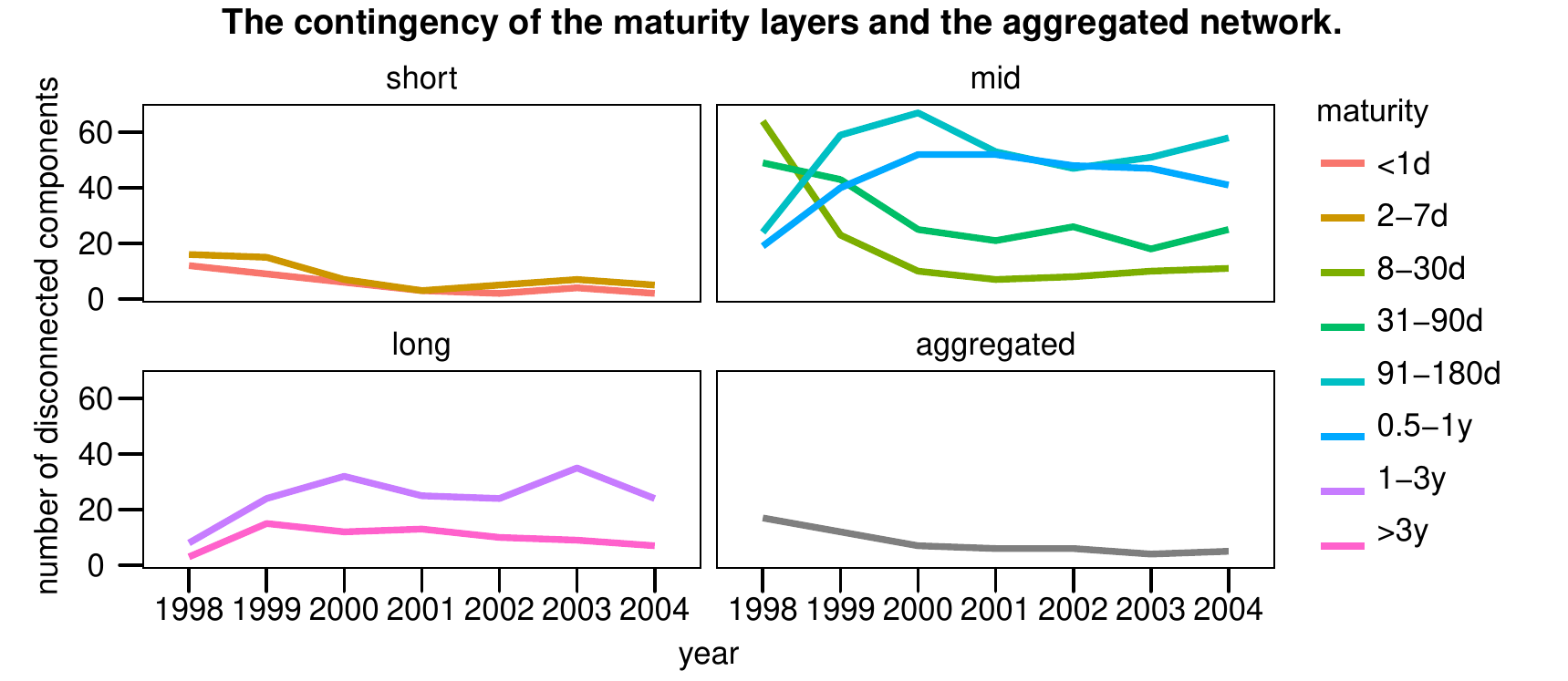}
\caption{To compare the number of (weakly) disconnected components with the number of active banks per year, consult Table~\ref{numbanks} on page~\pageref{numbanks}. The relative size of the largest weakly connected can be found in Fig.~\ref{clustering-gcsize}. \label{clustering-numcomp}}
\end{figure}

The remaining layers display the familiar pattern: the size of the largest connected component shrinks for longer maturities, while simultaneously the network becomes increasingly fragmented as the number of unconnected components increase, relative to the number of active banks (Fig.~\ref{clustering-numcomp}) -- both tend to suppress $D$.

\clearpage
\subsection*{Clustering coefficient}

Clustering coefficients measure to which extent banks form triangles. Put differently, they measure the tendency of connected nodes to have common neighbors in undirected views of interbank networks. 

According to Bargigli et al.~\cite{bargigli2015}, an inverse relationship between the degree and the clustering of a node is observed quite commonly. In the core-periphery picture, low clustering values of core nodes indicate that they essentially behave as star centers. A star graph exhibits zero clustering, as the periphery nodes are unconnected amongst themselves. Consequently, deviations from the star graph, which has idealized core-periphery structure, can be probed by measuring the local clustering coefficients of the periphery nodes. One can identify these heuristically by simply considering the nodes with low degree, so the aforementioned inverse relationship hints that the core-periphery structure may exhibit considerable complexity. A second and more robust implication is that the clustering coefficient of the complete interbank network is dominated by the clustering of low-degree nodes if this inverse relationship is observed together with heavy-tailed degree distributions.

High clustering has obvious implications for systemic risk~\cite{Vandermarliere2015} and therefore clustering coefficients are of interest to the interbank network literature. Table~1 in the main text shows, that at least a few studies claim opposite `typical' values for clustering coefficients. This may be due to the fact that the clustering tends to increase with longer time windows~\cite{bargigli2015}, or that the coefficients may not have been compared to the mean clustering coefficient obtained for random networks of the same size and number of edges.

The \emph{local} clustering coefficient~\cite{newman2003} of a node $i$ is defined in function of triangle motifs in undirected networks:
\begin{equation}
C_i = \frac{\text{number of of triangles connected to $i$}}{\text{number of triples on $i$}}, \label{local-C}
\end{equation}
where a triple on $i$ means an unordered pair of nodes connected via $i$, and possibly connected directly. If that is the case, the triple is counted as a triangle, ending up in the nominator of Eq.~\ref{local-C} as well. Thus $C_i$ expresses the degree of connectedness among the neighbors of $i$~\cite{Vandermarliere2012}. The (\emph{global}) clustering coefficient of an undirected $(N,E)$ network is then simply the average of the local coefficients:
\begin{equation}
C = \frac{1}{N} \sum_i C_i. \label{glob-C}
\end{equation}
To assess whether an observed network possesses a non-random clustering structure and thus a significant clustering coefficient $C_0$, we generate $n$ random networks of the same size, i.e. the same number of nodes and edges, having clustering coefficients $C_\text{rand} = \{C_1, C_2, \ldots, C_n\}$. Then we can calculate the $z$-score of the observed $C_0$ as
\begin{equation}
z = \frac{C_0 - \langle C_\text{rand} \rangle}{\text{sd}(C_\text{rand})}. \label{z-scores}
\end{equation}
Large $z$ indicates a significant value of $C_0$, the sign indicating more $(+)$ or less $(-)$ clustering than a random network of the same size.

\begin{figure}[h!]
\hspace*{-1cm}
\includegraphics{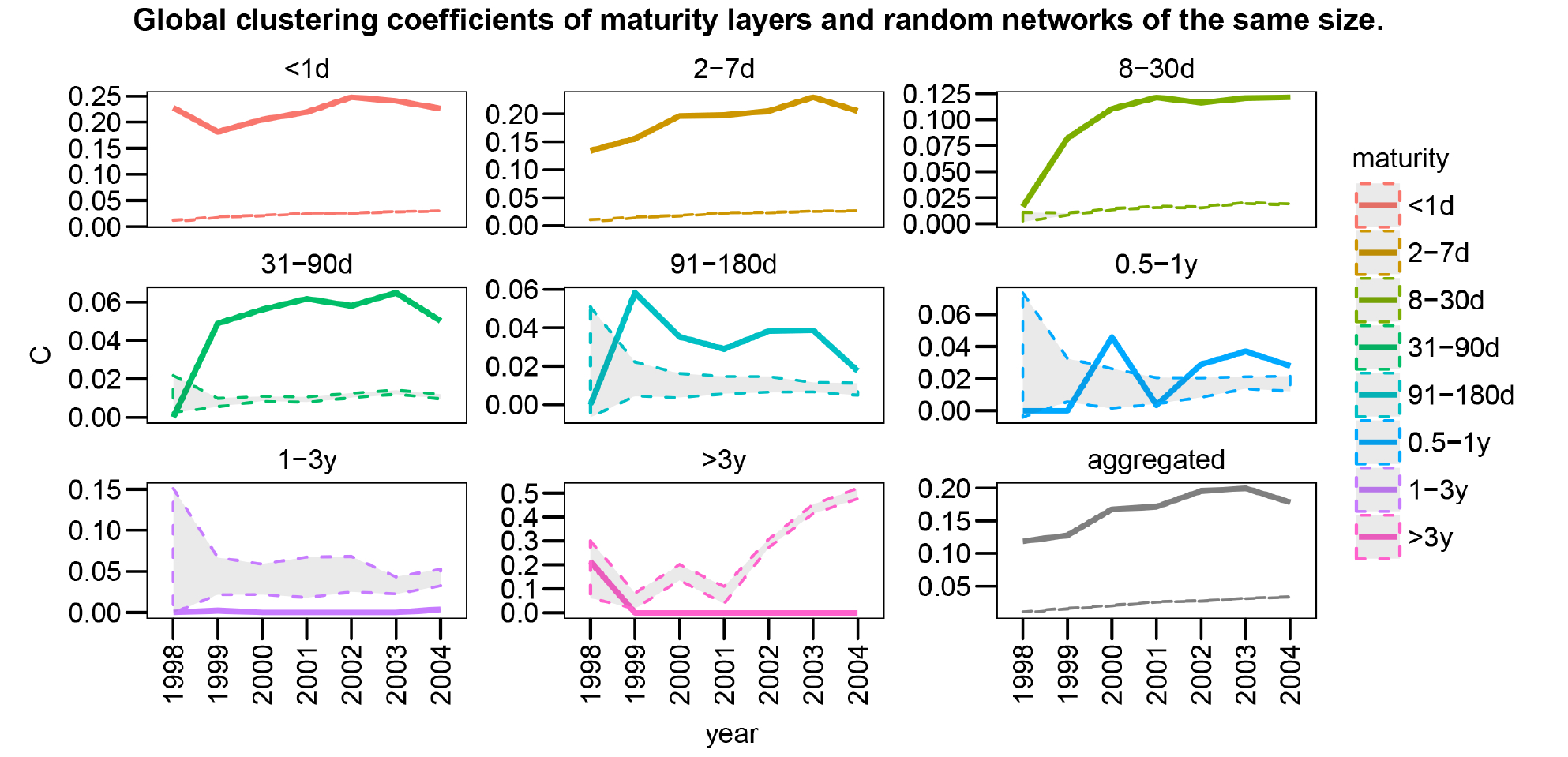}
\caption{Time series of clustering coefficients for the loan maturity layers and the aggregated network, together with a ribbon centered around $\langle C_\text{rand} \rangle$ of total width $2 \times \text{sd}(C_\text{rand})$. Following the methodology in \cite{bargigli2015}, a total of 100 Erdős–Rényi networks were generated to test the significance of the observed $C$'s. \label{clustering-ccoef}}
\end{figure}

We have calculated clustering coefficients and $z$-scores for the undirected view of the loan maturity layers using yearly time windows which are displayed in Fig.~\ref{clustering-ccoef}. A clear pattern can be seen: the clustering varies from significantly high to moderate (in order $G_{<1d}$, $G_{2-7d}$, and $G_{8-30d}$), via moderately low to insignificant (in order $G_{31-90d}$, $G_{91-180d}$, and $G_{0.5-1y}$), to significantly low ($G_{1-3y}$ and $G_{>3y}$). Since all layers exhibit heavy-tailed degree distributions, the sum in Eq.~\ref{glob-C} is dominated by the (local) clustering coefficients of banks with small degree. We can use $C$ to proxy the clustering of the periphery banks

Within that approximation, one sees that the core-periphery structure expected in the short maturity layers deviates considerably from the idealized star network, at least when looked at with yearly resolution. Notwithstanding the strong presence of intermediation found earlier, the lower-tier banks still trade extensively with each other, indicating that contagion risk is not located entirely in the high-tier banks.

In contrast, the two longest maturity layers, where source-sink structures are expected, exhibit considerable star-like structure, low degree nodes and periphery banks being equivalent in most cases. In addition to the low clustering, we recall the modest anticorrelation between in- and outdegree; these observations lead us to believe that $G_{1-3y}$ and $G_{>3y}$ behave as a sort of \emph{source-sink star layers}. These layers have star-like hub structure, but hardly any intermediation occurs in the hubs. They act simply as sources and sinks, generating and dissipating excess liquidity in the interbank network. Of course, one could argue that for loans with maturities of at least one year, no intermediation is possible within the scope of one year. Looking at Fig.~\ref{degrees-inout}, which completely aggregates time, however, we see that the banks with the largest degrees tend to be \emph{either} sinks or sources, especially in $G_{>3y}$. Thus the tendency of the hubs to be either sink or source, but not an intermediary, holds for time scales longer than one year. The next question is whether the star centers in the source-sink star layers are connected, or rather at the center of disconnected components. Disconnected components are plentiful compared to the number of active banks per year (see Fig.~\ref{clustering-numcomp} and Table~\ref{numbanks}), and less than half of the latter participate in the largest (weakly) connected component, as Fig.~\ref{clustering-gcsize} shows. Furthermore, we report that in the two source-sink star layers the average shortest path length is between 2 and 3 (see Fig.~\ref{distance-month}, which shows the average path length for $G_{1-3y}$ and $G_{>3y}$ on a \emph{monthly} basis as being typically 2), which, together with the proven existence of hubs, indicates a compact star-like structure. In a nutshell, the source-sink star layers $G_{1-3y}$ and $G_{>3y}$ are composed in general of many disconnected `islands', of which the largest exhibit an almost perfect star-like structure around a sink or source hub.

All results are highly significant, except for $G_{91-180d}$ and $G_{0.5-1y}$: these layers do not possess any structural clustering structure on their own. As always, the clustering in the aggregated network resembles mainly the first two layers. We further note that clustering is most volatile in times of crisis, and increases during the growing phase of the network~\cite{Vandermarliere2015}. 

\clearpage
\subsection*{Bank size mixing as degree size mixing}

Many studies point to \emph{disassortative mixing} with respect to the bank size, meaning that small banks trade mainly with large banks and vice versa. Given the core-periphery picture in these studies, one expects that the core banks are the large banks. Indeed, it is found that total bank assets are significant in explaining core membership~\cite{Craig2014}. As bank size correlates with total degree, we would also expect disassortative mixing of the banks with respect to the (total) degrees, i.e. high (low) degree nodes tend to be connected to low (high) degree nodes~\cite{newman2003}. In fact several studies report this as an additional stylized fact~\cite{aldasoro2016,bargigli2015}.

\begin{figure}[h!]
\includegraphics{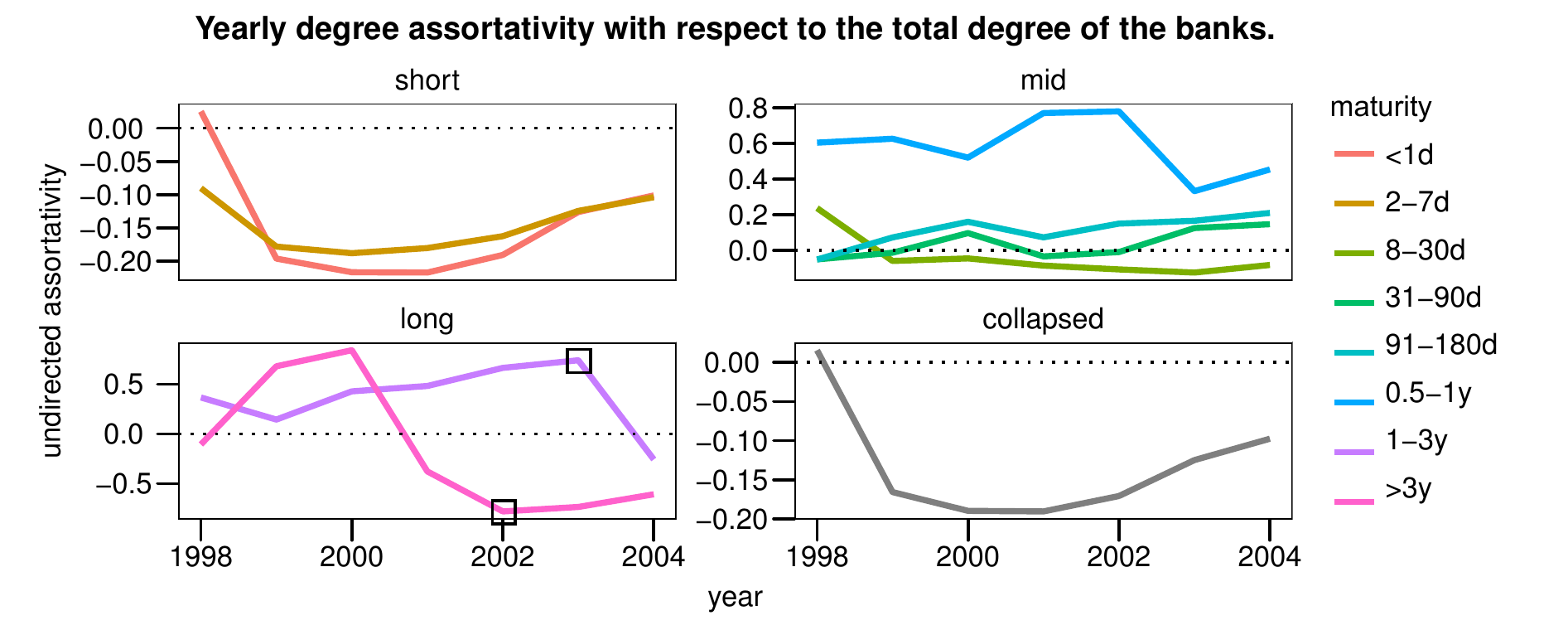}
\caption{The degree assortativity measures to which extent nodes with a given degree associate preferentially with other nodes of similar degree -- see Equation (2) in \cite{newman2003as} for the formal definition. In this case the type of degree considered is the total degree, i.e. the sum of the in- and outdegree. A yearly resolution was chosen because monthly aggregation suffered from bad statistics starting from $G_{91-180d}$. \label{disassortativity-yearly}}
\end{figure}

\begin{figure}[h!]
\includegraphics[width=\textwidth]{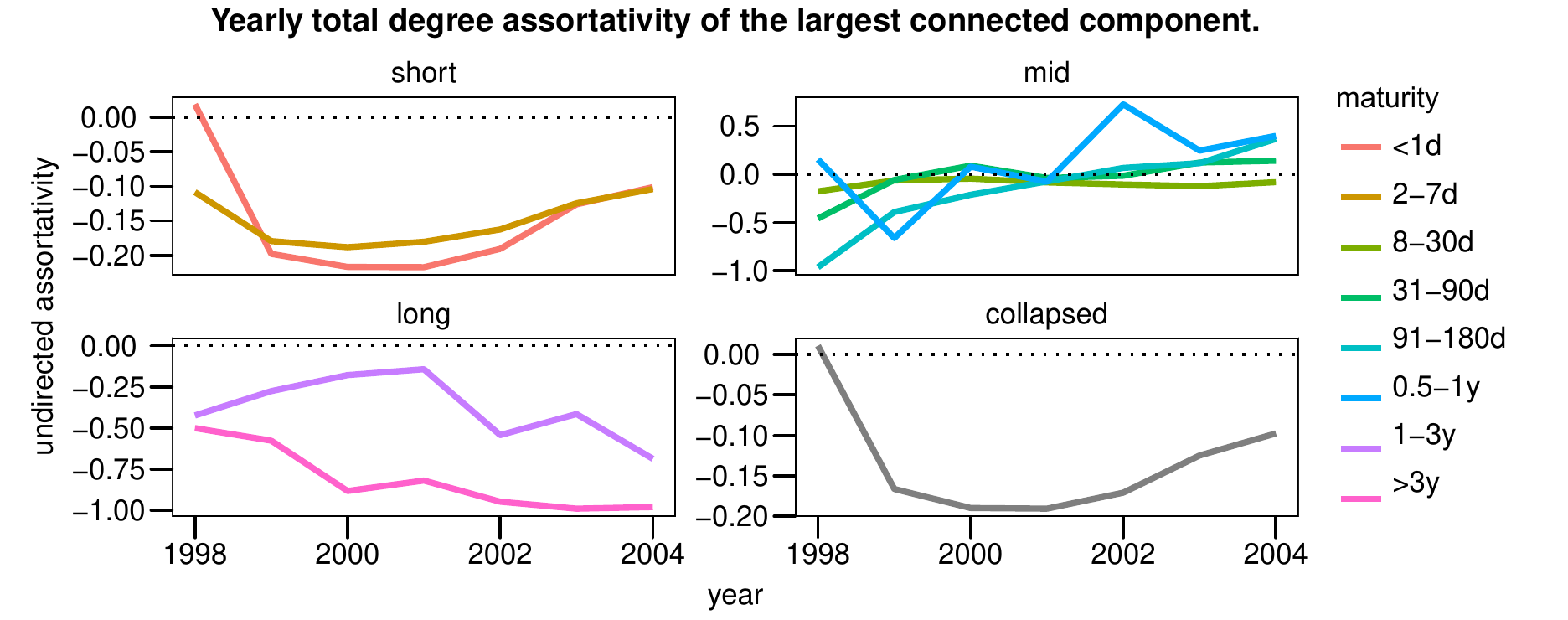}
\caption{The degree assortativity for the total degrees throughout the years. The degree assortativity is also called degree correlation \cite{newman2003}. \label{disassortativity-lcc}}
\end{figure}

Figure~\ref{disassortativity-yearly} shows the assortativity with respect to the total node degrees on yearly basis, which is also called the total degree correlation. The layers with short maturity show a clear preference for the low-degree nodes to attach to the high-degree ones. This preference weakens when we look at longer maturity layers. The assortativity or the long maturity layers is pushed up by the large number of disconnected clusters, in most cases simply isolated pairs of trading banks. The disassortative degree mixing in the larger clusters is present, caused by their star-like structure. As the number of unconnected components grows drastically, we would expect that the same mechanism is behind the high assortativity for $G_{91-180d}$ and especially for $G_{0.5-1y}$. Interestingly, this expectation turns out to be false: evaluating the degree assortativity only in the largest connected component results in lower values for all layers (notably in those where star-like structures were expected based on the clustering coefficient), but it remains strongly positive for $G_{91-180d}$ and $G_{0.5-1y}$ during the stable phase of the network (Fig.~\ref{disassortativity-lcc}).

\subsection*{Summary}
A summary of the results of this maturity layer analysis can be found in Table 1 in the main text. While certain features here are indicative of known structures in the interbank lending network literature, we refrain from making any such claims. Rather we report the findings here as is and only make claims about function and structure in the main text where steps have been taken to gather statistical evidence for both the construction of the layers as the occurrence of structure in such layers.
\clearpage

\section{OG inference with non-contiguous binning}

\begin{figure}[h] 
\centering
\includegraphics{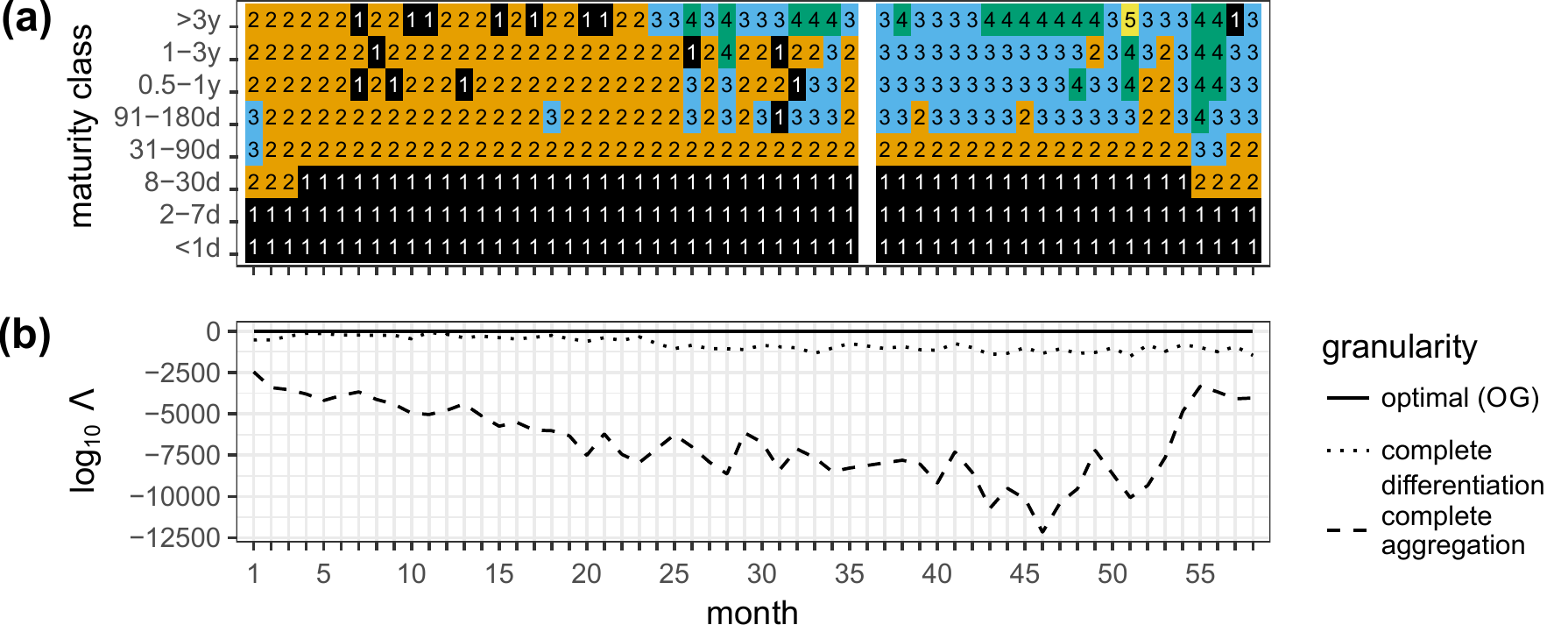}
\caption{Results of the layered SBM with non-contiguous binning. The optimal granularity (OG) with respect to the loan maturity classes for the Russian interbank lending network. The OG corresponds to the granularity of the best-fit layered SBM. 
\textbf{(a)} Monthly time series of the OG inferred from the monthly interbank lending network. Each OG bin (OGB) holds one or more maturity classes and is labelled by an OGB index $1,2,3,4$ and indicated by a colour. The OGB index runs according to the shortest maturity class in the bin. The bin with the shortest maturity class in it is numbered OGB 1, etc. The OGBs correspond to lending patterns between the bank groups that differ from each other in a statistically significant way.
\textbf{(b)} Temporal evolution of the $\log_{10}$ of the posterior odds ratio $\Lambda$ (for its definition see Eq.~4 in the main text) of the layered SBM for three different granularities: (i) the OG; (ii) complete loan maturity differentiation; (iii) complete loan maturity aggregation. Note that the OG inferred from the algorithm is consistently preferred. Complete aggregation is decisively rejected as an optimal representation of the mesoscale of the monthly interbank lending network \label{non-cont-olp}}.
\end{figure}

Here we present the results of the OG inference where non-contiguous binning was allowed in Fig.~\ref{non-cont-olp} (as opposed to the contiguous binning in the main text). This allows for partitions where short maturity loans are merged with long maturity loans. As explained in the main text, the creation of a new bin in the OG needs to be warranted by enough statistical evidence for it to be necessary; i.e. that it will contain a lending pattern that is significantly different from the lending patterns in the other bins. In those cases where very long maturity loans are merged with short maturity loans, we suspect that there is hardly any actual structure (and thus not enough statistical weight) in the long maturity layer due to its sparcity. It then makes sense that such layer is merged with the densest layer available since this merging will cause the least amount of new connections to be introduced relative to the other, less dense, bins.

Comparing Fig.3 (a) and (b) in the main text to Fig.~\ref{non-cont-olp} (a) and (b) shows the results to be qualitatively the same.

\clearpage


\section{Number of active banks per group}

\begin{figure}[h] 
\centering
\includegraphics{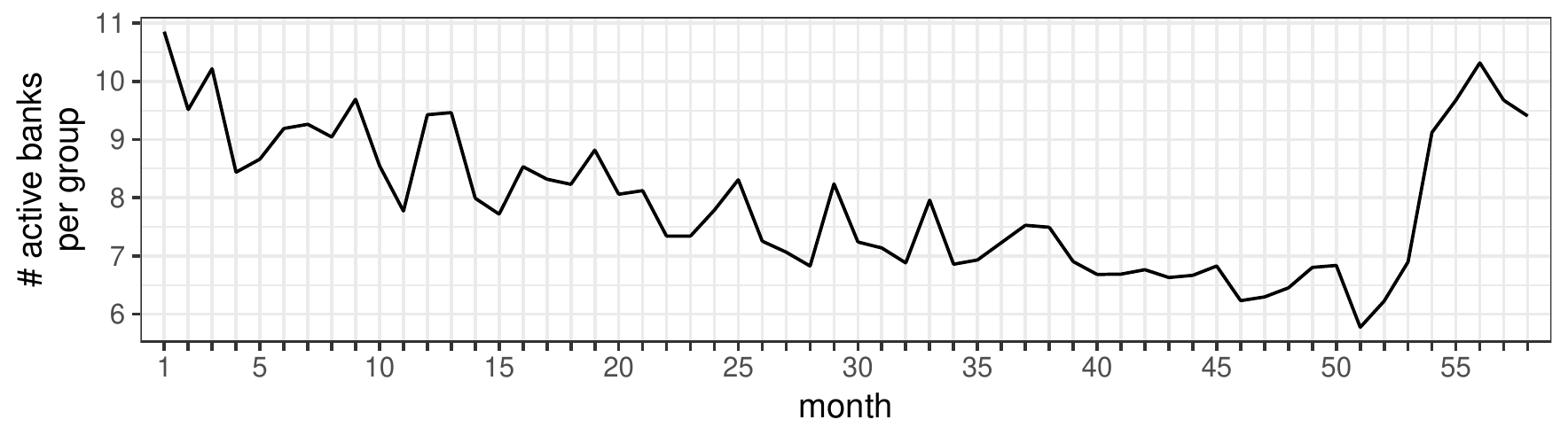}
\caption{The number of active banks per group is obtained by dividing the number of active banks by $B$, i.e. the number of groups in the inferred bank groups $\{b_i\}$.}
\end{figure}
\clearpage

\section{Activity characteristics per OGB}

\begin{figure}[h] 
\centering
\includegraphics{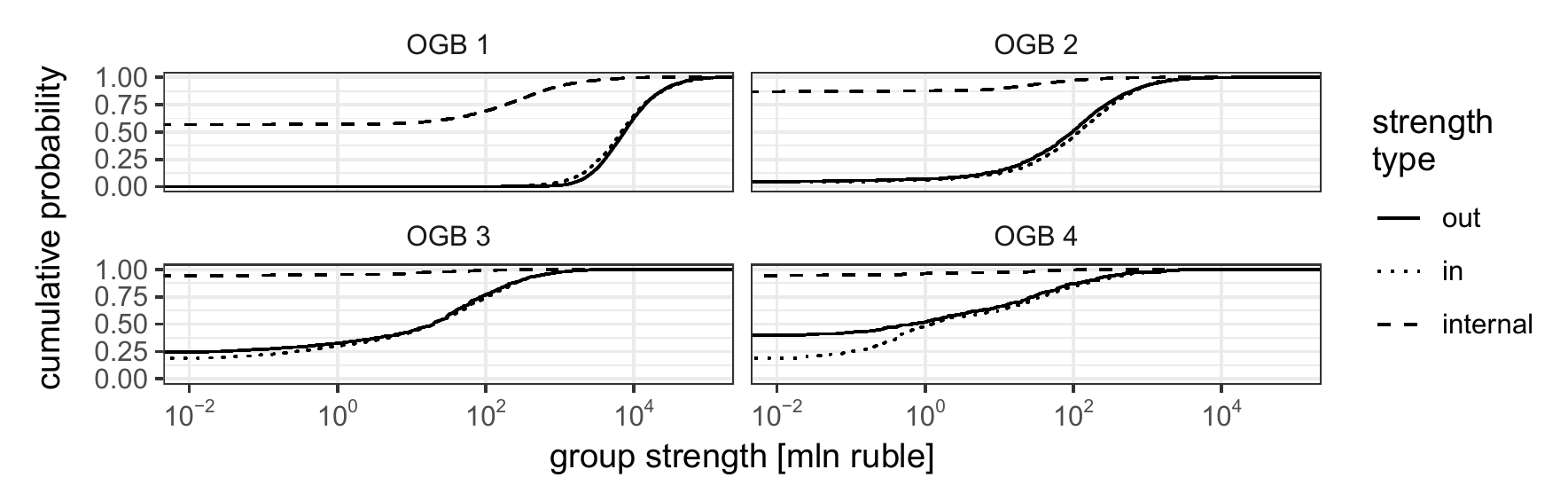}
\caption{The time-integrated cumulative density functions (CDFs) of the group instrengths, outstrengths and internal strengths for each OGB. Zero group strengths are represented by non-vanishing CDFs at the origin.\label{strength}}
\end{figure}

We investigate the OGBs and their typical characteristics. Fig.~\ref{strength} shows the distribution of the log group strengths across time and per OGB. We discern three types of group strength: outstrength, instrength and internal strength. The group outstrength in a certain month of bank group $b$ is the total amount lent by all banks belonging to $b$ to banks belonging to another group $b' \neq b$. In symbols,
\begin{equation}
    s_b^\text{out} = \sum_l \sum_{i,j,k} x_{ijk}^l \delta_{b_i,b} (1 - \delta_{b_j,b})
\end{equation}
The two other strength types and the constraint that the maturity layer $l$ must be in a given OGB are obtained likewise from this definition.

We see that the groups for OGBs 2, 3, 4 show definite ``anti-community characteristics'', i.e. they hardly lend or borrow internally. There is also an interesting asymmetry between group instrength and group outstrength in OGB 4. Looking at the group strength modes, each OGB is seen to function on its own magnitude scale (see Fig. 5 in the main text). 
\clearpage

\section{Other percentiles for instrength/strength of important banks}
\begin{figure}[h] 
\centering
\includegraphics{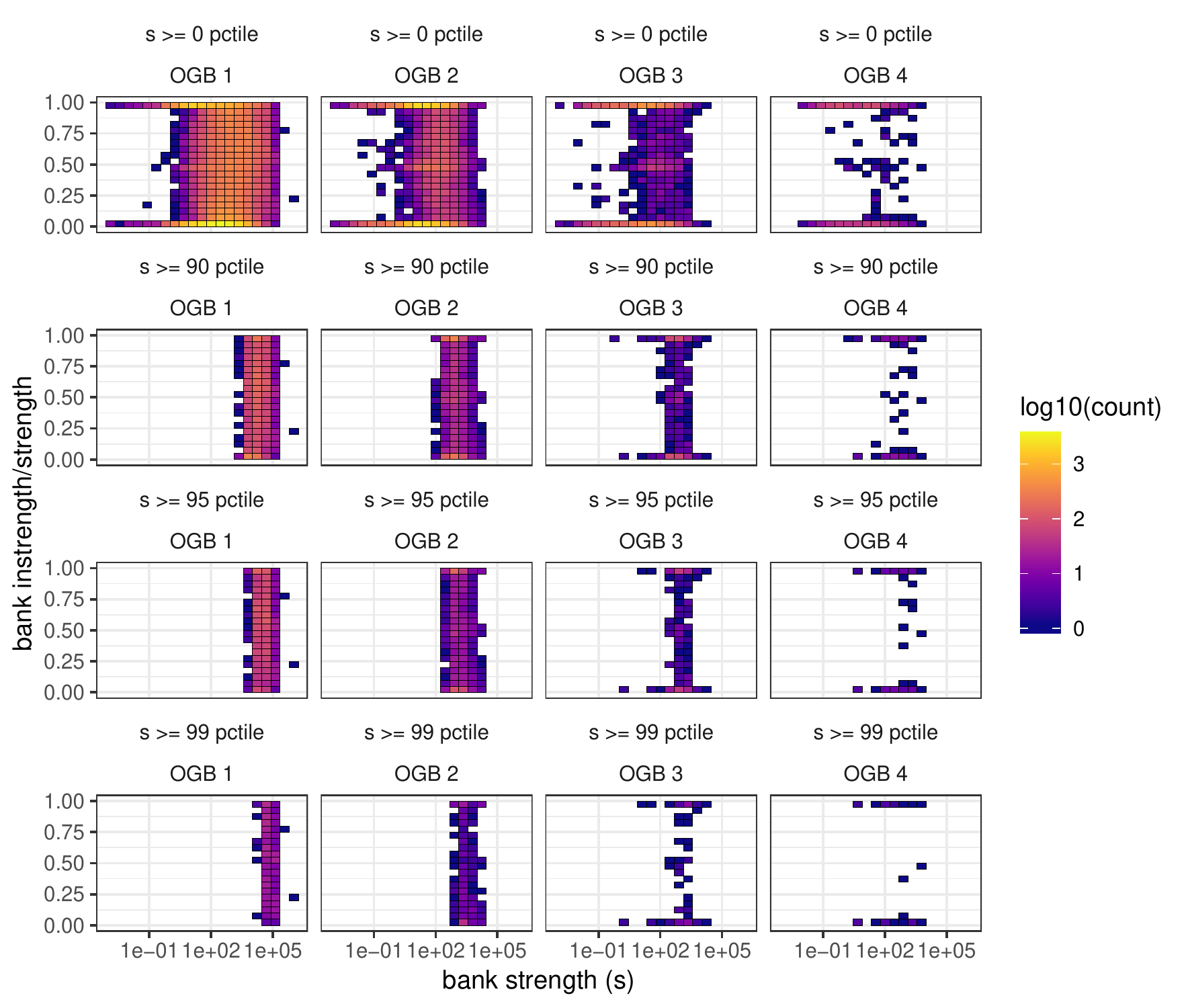}
\caption{The time-integrated distribution of the strength and the instrength/strength ratio of the monthly ``important banks'' for the four OGBs. As in Fig.~4(b) in the main text we gauge a bank's importance by its strength: For a given month a bank is deemed important if its strength lies in the top $X$, with $X$ varying per row of subplots (the top $X$\% is defined by the ($1-X$)th percentile). With growing OGB index, the important banks increasingly tend to either lend or borrow. Banks in OGB 1 and OGB 2 tend to balance lending and borrowing. This suggests that the economic function of the important banks in the various OGBs changes from financial intermediation (short maturity bins) to financing (long maturity bins) on a monthly time scale.\label{instrength}}
\end{figure}

We look at the lending behaviour of individual banks. As in the main text we again use the bank strength to single out the ``important'' banks. We show that the conclusions from the main text hold for different bank strength cutoffs. Each row of subplots in Fig.~\ref{instrength} uses a different cutoff, respectively with the bank strength ($s$) unlimited, limited to the top 10\%, top 5\%, and the top 1\%. At monthly time scales the important banks in ``short'' bins tend to both lend and borrow equal amounts of money, while the important banks in the ``long'' bins tend to either lend or borrow.

\clearpage

\section{The trust crisis}
\label{app:crisis_timeline}
The trust crises consist of the announcement crisis in the second half of 2003 and the summer of 2004 crisis~\cite{Karas2012, Porfiriev, vandewiel2013,VandenHeuvel2015}.
\newline
\newline
\begin{centering}
\hspace{-60pt}
\begin{tabular}{rl}
end of July 2003 &%
\begin{minipage}[t]{0.65\columnwidth}%
Start of announcement crisis. Decline in reciprocal bank interactions.%
\vspace{1ex}
\end{minipage}\tabularnewline

September 2003 &%
\begin{minipage}[t]{0.65\columnwidth}%
The network starts to fall apart because of the distrust among banks.
\vspace{1ex}
\end{minipage}\tabularnewline

December 2003 - January 2004 &%
\begin{minipage}[t]{0.65\columnwidth}%
Back to normal liquidity.
\vspace{1ex}
\end{minipage}\tabularnewline

March &%
\begin{minipage}[t]{0.65\columnwidth}%
Peak of interbank lending market recovery from early trust crisis. The bank lending reciprocity, however, does not regain pre-crisis levels.%
\vspace{1ex}
\end{minipage}\tabularnewline

\small{following weeks} &%
\begin{minipage}[t]{0.65\columnwidth}%
Three stages of the summer of 2004 crisis.%
\vspace{1ex}
\end{minipage}\tabularnewline

First stage: April &%
\begin{minipage}[t]{0.65\columnwidth}%
Volatility on the interbank market with even higher lending rates than later stages but without significant outflow of individual's deposits. Demand for liquidity was caused by policy changes and statements of the CBR; no perception of crisis by the banks themselves, yet this financial instability undoubtedly impacted the crisis to come.%
\vspace{1ex}
\end{minipage}\tabularnewline

Second stage: May &%
\begin{minipage}[t]{0.65\columnwidth}%
CBR deprives Sodbusinessbank of its license.%
\vspace{1ex}
\end{minipage}\tabularnewline

May 19-21 &%
\begin{minipage}[t]{0.65\columnwidth}%
Several conflicting statements from authorities about the deposit insurance for the clients of Sodbusinessbank make the depositors increasingly uneasy.%
\vspace{1ex}
\end{minipage}\tabularnewline

Third stage: June 3 &%
\begin{minipage}[t]{0.65\columnwidth}%
Crisis now definitively developed beyond Sodbusinessbank alone. Banks start to introduce additional control measures; several suspend lending activities on the interbank market.%
\vspace{1ex}
\end{minipage}\tabularnewline

June 11 &%
\begin{minipage}[t]{0.65\columnwidth}%
CBR changes policy rates to accommodate the banks, officially for the low liquidity on the interbank short-term market.%
\vspace{1ex}
\end{minipage}\tabularnewline

June 21-22 &%
\begin{minipage}[t]{0.65\columnwidth}%
Peak of the crisis.%
\vspace{1ex}
\end{minipage}\tabularnewline

July 13 &%
\begin{minipage}[t]{0.65\columnwidth}%
Interbank market starts to stabilize.%
\vspace{1ex}
\end{minipage}\tabularnewline

July 16 &%
\begin{minipage}[t]{0.65\columnwidth}%
The `crisis of confidence' is declared to be at an end.%
\vspace{1ex}
\end{minipage}

\end{tabular}
\end{centering} \\

\hl{Investigations of money laundering led the CBR to deprive Sodbusinessbank of its license in May 2004. The following mutual suspicion led to a drying up of liquidity on the interbank market in the summer of 2004. Roughly one year earlier, this investigation was announced, which caused a smaller trust crisis (i.e. the announcement crisis). In May 2004, a crisis on the Russian interbank market was triggered. In particular, on May 13 the Central Bank of Russia recalled the license from Sodbiznesbank on accusations of money laundering and sponsorship of terrorism. It was the first bank to be closed on these grounds in Russia. This unexpected closure caused panic on the interbank market since banks suspected other banks would follow suit, but they had no reliable information on who these banks might be.}

\hl{On May 16 the head of Federal Financial Monitoring Service, Viktor Zubkov, announced publicly that his agency suspected another ten banks of similar violations. Because this announcement was not accompanied by the actual list of suspected banks, it just fuelled the hysteria on the interbank market. Rumours about the identities of these ten banks started to spread rapidly. Soon several inconsistent blacklists were circulating in the banking community as bankers tried to guess which banks were officially suspected of money laundering. Anecdotal evidence suggests that banks were actively helping to spread the rumour by removing themselves from the list and adding competitors in an attempt to escape the carnage. By consequence, The union of all the blacklists circulating in the banking community expanded in a few days to include dozens of banks, including several market leaders.}

\hl{In the presence of total uncertainty about the quality of their counterparties, banks began to reduce limits on each other, which reverberated into an acute liquidity drought on the interbank market~\cite{Paranyushkin2009RussianProblems}. The turnover volume on the interbank market dropped spectacularly.
Later the original source of the panic, Viktor Zubkov, shockingly announced that ``the Federal Financial Monitoring Service has no blacklist''. The deputy Minister of Internal Affairs similarly announced that ``the Interior Ministry has no such list. We have no plans to persecute any banks.''For further descriptions (in Russian) of this episode in Russian history see~\cite{Simonov2009Banks1988,Boyarskiy2018Modern1988}. Up till now it remains unclear whether an official list actually existed. Important for us is that the 2004 meltdown on the Russian interbank market was based on rumours and unrelated to shocks to the fundamentals of Russian banks or the Russian economy at large.}

\hl{This non-fundamental and exogenous mutual trust crisis provides a great example of what structural changes occur during a shock to the non-fundamentals of the ILN. The advantage of the trust crisis compared to the $2007$-$2009$ financial crisis is that, while the financial crisis was caused elsewhere and only indirectly found its way into Russia (via oil-prices, etc.), the trust crisis had its epicenter inside of Russia.}


\clearpage

\section{Maturity classes in the ILN literature}

An overview of the ILN literature using multilayer approaches and how their granularity compares to the current dataset.

\begin{table}[h]
\caption{Classifications of loan maturity in terms of our maturity classes used by the Central Bank of Russia and several studies that have included multilayered ILNs.\label{mux-classes}}
\centering
\begin{tabular}{@{}lccllllcl@{}}
\toprule
                                                  & \multicolumn{1}{l}{<1d}   & \multicolumn{1}{l}{2-7d} & 8-30d & 31-90d & 91-180d & 0.5-1y & \multicolumn{1}{l}{1-3y}     & >3y    \\ \midrule
                                                  Central Bank of Russia \cite{guleva2015}          & \multicolumn{1}{|c}{instant}   & \multicolumn{2}{|c|}{short term}                   & \multicolumn{5}{|c|}{long term}                                                     \\
Bargigli et al. (2015) \cite{bargigli2015} & \multicolumn{1}{|c}{overnight} & \multicolumn{5}{|c|}{short term}                                                                       & \multicolumn{2}{|c|}{long term} \\
Aldasoro and Alves (2016) \cite{aldasoro2016}            & \multicolumn{6}{|c|}{short term}                                                                                   & \multicolumn{2}{|c|}{long term}                         \\
Montagna and Kok (2016) \cite{montagna2016multi}              & \multicolumn{4}{|c|}{short term}                                                & \multicolumn{4}{|c|}{long term}                                                            \\
\bottomrule
\end{tabular}
\end{table} 
\clearpage

\section{Term structure of interest rates and loan volumes in the interbank loan market}\label{section-term-structure}

\begin{figure}[h] 
\centering
\includegraphics{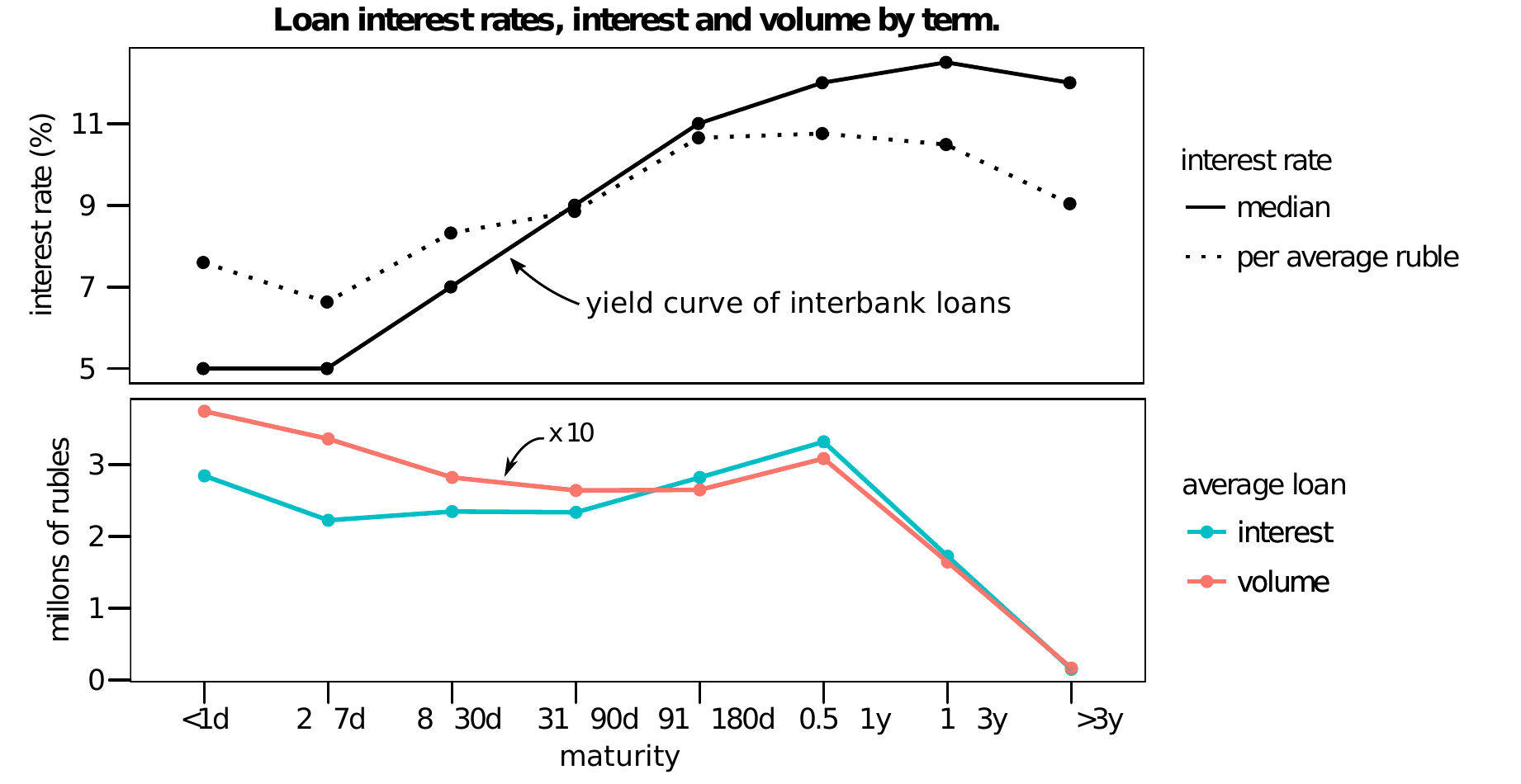}
\caption{\hl{(top panel) The global categorical yield curve is the median interest rate over time per term. An alternative characterization is the interest rate per average ruble lent/borrowed, defined as the average interest weighed by the loan volumes. This statistic is robust to whether one excludes or includes the interest rate outliers. (lower panel) The average loan interest received/paid per term in millions of rubles and average loan volume (size) lent/borrowed in tens of millions of rubles.}\label{term-summary}}
\end{figure}

\subsection*{\hl{The global categorical yield curve}}

\hl{The yield curve is obtained after ``averaging'' interest rates $r$ over time and by maturity class. The averaging technique chosen is the median due to extreme outliers which disproportionately affect the mean: the standard deviation of the complete $r$ population (including outliers) is $\sigma \approx 40\%$, yet $99$\% of the population is contained in $[0,\sigma]$. Within that interval the standard deviation is about $7$\%, which is a more sensible measure of the interest rate dispersion. By using the median, we avoid choosing any cutoff.}

\hl{Yield curves are usually considered with \emph{continuous} maturities ranging from one day to several years, but the available data only records maturity \emph{classes}. Although methods exist to estimate the continuous yield curve from discrete data~\cite{anatolyev2003}, we will only consider interest rates (and interests etc.) per category, i.e. per maturity class.}

\hl{The solid line in the top panel of Figure~\ref{term-summary} exhibits a typical stylized fact of the yield curve. It is upward sloping and has a convex shape, except for the >3y term, i.e. maturities three years or longer. The unusual steepness of the curve in Figure \ref{term-summary} is an artifact caused by the term categories; as they progress, they bucket a growing amount of maturities, so a continuous yield curve would be horizontally stretched with respect to the categorical yield curve. The upward slope is usually explained by classical \emph{expectations theory} (ET). According to ET, interbank lending rates dynamics are determined by the structure of liquidity supply and demand \cite{Egorov2013}; the long-term interest rate is an average of expected future short-term rates, plus a term premium that increases with longer terms to compensate risk-averse lenders for the interest risk, which arises for lenders from fluctuating interest rates with respect to the base deposit policy rate. One can also extend ET by including separate premiums for liquidity risk (selling loans on the secondary market tends to be harder as their maturities lengthen) and default risk, also called counterparty risk, which is in theory governed by the credit rating of the borrower. Normally the short-end (long-end) of the yield curve is dominated by the liquidity (default) risk; both are considered components of interbank lending risk~\cite{tabak2001,russell1992}. Each risk contributes to the upward slope of the yield curve given normal market conditions.}

\hl{The interest per average ruble is also displayed in the top panel of Figure \ref{term-summary}. Loans with longer maturities are the most profitable for lenders. The drop for the >3y term could point to the fact that these contracts may be made on more amicable (flexible) terms; this would imply lower perceived interbank risk which could explain the negative slope between the 1-3y and >3y categories.} 

\begin{figure}[h] 
\centering
\includegraphics{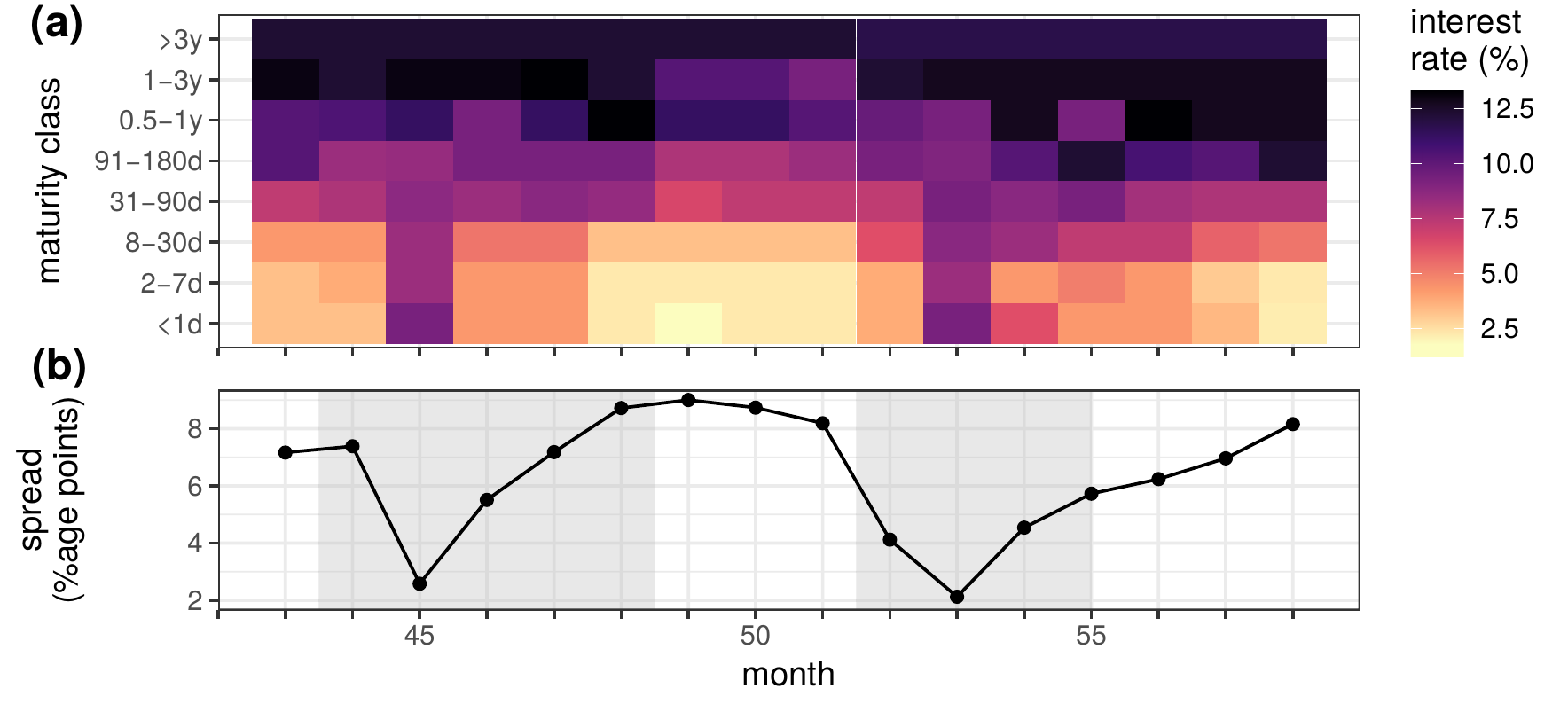}
\caption{\hl{Monthly categorical yield curves (i.e. aggregated by maturity class; top panel) together with the spread (bottom panel) from July 2003 (month up until November 2004. This period is embedded in the emergent maturity phase of the network and includes the announcement and summer of 2004 crises (these are the trust crises -- see Section~\ref{app:crisis_timeline}), both shaded in grey on the bottom panel. The first shaded region spans the period from August 1st 2003 until January 1st 2004. The second shaded region spans the April 1st 2004 until July 16th 2004. The CBR lowered the overnight rate from 16 to 14\% on January 15, 2004 and again to 13\% on June 15, 2004. The yield curve is the median interest rate per maturity class, as in Figures~\ref{term-summary}; the spread is defined in the text.}\label{yield-curve-movement}}
\end{figure}

\hl{To illustrate the dynamics that underlie the global yield curve, Figure~\ref{yield-curve-movement}(a) plots the categorical yield curve for each of the last 16 months in the data. In this period the by now mature interbank network deals with the trust crises which are explained in more detail in Appendix~\ref{app:crisis_timeline}. Figure~\ref{yield-curve-movement}(b) contains the \emph{(yield) spread}, which we define as the difference in average yield between "long" maturities and "short" maturities. We define these maturities as two classes: the "long maturity" class consists of maturity classes (\mty{1-3y}, \mty{>3y}) and the "short maturity" class consists of maturity classes (\mty{<1d}, \mty{2-7d}, \mty{8-30d}). The general shape of the curve is not sensitive to the definition of the "long maturity" class; the "short maturity" class is motivated in the paper, while the "long maturity class" is based on Basel recommendations and is also used in ~\cite{bargigli2015,aldasoro2016}.}

\hl{In normal times, interbank markets are among the most liquid in the financial sector: banks prefer to lend out excess cash since the central bank's interest rate on excess reserves is smaller than rates available in interbank markets~\cite{heider2009}. During trust crises, the perceived default risk grows, which inflates interest rates according to ET. Riskier banks, i.e. banks at risk of being in financial distress, exert an externality on safer banks who subsidize their liquidity\cite{heider2009}. If the crisis gets worse, this externality on safer banks is so costly that they leave the unsecured market, and liquidity rich banks may prefer to hoard liquidity instead of lending it out to an adverse selection of borrowers; the interbank lending market dries up.}

\hl{We see that this mechanism is indeed captured by the spread curve in the bottom panel of Figure~\ref{yield-curve-movement}: low spread seems to indicate abnormal market conditions with low liquidity because the short term interest rates increase quite relatively fast. Note in the upper panel that the long term interest rates stay almost constant during the crises and drop slightly during the recovery in between (roughly from January until April 2004). This suggests that we cannot make an analogy with the typical \emph{inverted yield curves} of e.g. treasury securities that are associated with (predicting) recessions. Indeed, according to ET these are yield curves with negative slopes because investors have poor expectations of future interest rates. In contrast, we see for our data that the short term interest rates rise quickly during the trust crisis while the long term rates hardly change.}

\subsection*{\hl{Characteristics of loan volumes and interests}}

\hl{The lower panel of Figure~\ref{term-summary} shows the average interest and volume per term. The loan volumes are log-normally distributed~\cite{Vandermarliere2015} and this holds remarkably well for the interest too, especially for shorter terms. Because of the considerable variance on a linear scale, these averages may be interpreted only as rough order of magnitude estimates. With this in mind, Figure~\ref{term-summary} shows that interest rate and volume are roughly negatively correlated: except for the bump at 0.5-1y and the case >3y, the volumes decrease as the interest rates increase. This can also be seen in the slowly varying average interest for the first five term categories.}

\hl{Table~\ref{term-volume} lists the total loan volumes by term and per year. As stated there (p.~\pageref{term-volume}), we observe that the relative importance of each term segment, as measured by the total volume of loans traded within it, follows the ranking of the loan terms. This, together with the typical volumes and interest rates in Figure~\ref{term-summary} and the lending activity in Table~\ref{numloans} (p.~\pageref{numloans}), supports the conclusion that the Russian interbank network exhibits a distinct hierarchy with respect to the loan maturities, which we could summarize by saying that \emph{banks lend greater volumes at lower interest rates more often for shorter loan terms.}} 
\clearpage
\renewcommand{\refname}{Appendix references}

\end{document}